\documentclass[manuscript,screen]{acmart} % for arXiv, no line number

\setcopyright{acmcopyright}
\copyrightyear{2022}
\acmYear{2022}
\acmDOI{}

\usepackage[utf8]{inputenc} % allow utf-8 input
\usepackage[T1]{fontenc}    % use 8-bit T1 fonts
\usepackage{hyperref}       % hyperlinks
\usepackage{url}            % simple URL typesetting
\usepackage{booktabs}       % professional-quality tables
\usepackage{amsfonts}       % blackboard math symbols
\usepackage{nicefrac}       % compact symbols for 1/2, etc.
\usepackage{microtype, float, wrapfig}      % microtypography
\usepackage{xcolor} 
\usepackage{bbm}
\usepackage[figuresright]{rotating}
\usepackage{amsbsy,amsmath,amsthm,latexsym,multicol,multirow,epsfig,color,subfigure,setspace,array,mathrsfs,bm,algpseudocode,algorithm}

\usepackage{graphicx}
\usepackage{enumitem}

\usepackage{natbib}

\newcounter{assumption}[section]

\newcounter{remark}[section]

\newcounter{example}[section]

\newcommand{\x}{\bm{x}}

\newcommand{\RR}{\mathbb R}

\newcommand{\bea}{\begin{eqnarray}}
\newcommand{\eea}{\end{eqnarray}}
\newcommand{\beaa}{\begin{eqnarray*}}
\newcommand{\eeaa}{\end{eqnarray*}}

\begin{document}

\title{CPS Attack Detection under Limited Local Information in Cyber Security: A Multi-node Multi-class Classification Ensemble Approach}


\author{Junyi Liu}
\authornote{Both authors contributed equally to this research.}
\email{liujunyi20@mails.tsinghua.edu.cn}
\affiliation{%
  \institution{Weiyang College, Tsinghua University}
  \city{Beijing}
  \country{China}
}

\author{Yifu Tang}
\authornotemark[1]
\email{tangyf20@icloud.com}
\affiliation{%
  \institution{Department of Mathematics, Zhili College, Tsinghua University}
  \city{Beijing}
  \country{China}
}

\author{Haimeng Zhao}
\affiliation{%
  \institution{Zhili College, Tsinghua University}
  \city{Beijing}
  \country{China}
}
\email{haimengzhao@icloud.com}

\author{Xieheng Wang}
\affiliation{%
  \institution{Tsinghua University}
  \city{Beijing}
  \country{China}
}
\email{wxh19@mails.tsinghua.edu.cn}

\author{Fangyu Li}
\authornote{Corresponding author.}
\affiliation{%
 \institution{Beijing University of Technology}
 \city{Beijing}
  \country{China}}
\email{fangyu.li@bjut.edu.cn}

\author{Jingyi Zhang}
\authornotemark[2]
\affiliation{%
  \institution{Department of Industrial Engineering, Center for Statistical Science, Tsinghua University}
  \streetaddress{30 Shuangqing Rd}
  \city{Beijing}
  \country{China}}
\email{jingyizhang@tsinghua.edu.cn}

%\renewcommand{\shortauthors}{Liu et al.}

\begin{abstract}

Cybersecurity breaches are the common anomalies for distributed cyber-physical systems (CPS). However, the cyber security breach classification is still a difficult problem, even using cutting-edge artificial intelligence (AI) approaches. In this paper, we study the multi-class classification problem in cyber security for attack detection. A challenging multi-node data-censoring case is considered. In such a case, data within each data center/node cannot be shared while the local data is incomplete. Particularly, local nodes contain only a part of the multiple classes. In order to train a global multi-class classifier without sharing the raw data across all nodes, the main result of our study is designing a multi-node multi-class classification ensemble approach. By gathering the estimated parameters of the binary classifiers and data densities from each local node, the missing information for each local node is completed to build the global multi-class classifier. Numerical experiments are given to validate the effectiveness of the proposed approach under the multi-node data-censoring case. Under such a case, we even show the out-performance of the proposed approach over the full-data approach.
\end{abstract}




\maketitle

\section{Introduction}
% for IEEE journal papers produced under \LaTeX\ using
% IEEEtran.cls version 1.8a and later.
% % You must have at least 2 lines in the paragraph with the drop letter
% % (should never be an issue)
% I wish you the best of success.


% \hfill September 17, 2014

%\IEEEPARstart{W}{ith} 
Over the past few decades, collecting and storing data in multiple local data centers/nodes rather than one single node has become more common in various areas due to the fast-developing technologies in data collection and processing \cite{kim2014long,ye2020cyber,li2019online,jiang2020prevalence,rieke2020future,Li2020review,zeng2021simple,guo2020systematic}. Such multi-node data then brings challenging environments in data analysis and integration \cite{Li2020review,Integration}, and has a higher risk of being attacked by hackers \cite{ding2018survey, khraisat2021critical}. 
As a result, cyber security problems have gained more attention in many fields, such as the autonomous driving system \cite{arnold2019survey,feng2020deep}, the AI-assisted diagnosis and treatment system \cite{jin2020ai,mitsala2021artificial}, and the Internet of Things (IoT) system \cite{thamilarasu2019towards,smys2020hybrid,khraisat2021critical}. %Such systems apply the state-of-art techniques developed in machine learning to complete certain tasks and compensate for system failures self-adaptively and automatically \cite{antsaklis1989towards}.
In cyber security problems, attack detection is of particular interest \cite{ding2018survey, khraisat2021critical}. There are a large number of works done for attack detection in the field of cyber security. Aitor Belenguer et al. \cite{Review} review applications of Federated Learning in Intrusion Detection. They introduce the development of intrusion detection systems and works including centralized learning, federated learning, and FEDAVG algorithm. They discuss the limitation of current works and future technologies. Many of the reviewed works associate the intrusion detection problem with the classification problem.
Besides, \cite{Comparative} compare various machine learning  methods for intrusion detection systems, including supervised learning, unsupervised learning, and reinforcement learning. They discuss both binary classification and multi-class classification scenarios according to the IoT-23 dataset. Classification problems play an important role in intrusion detection systems since different types of attacks could be seen as different classes. And among all the aforementioned attack-detection approaches, one of the main tasks is multi-class classification. 

\begin{figure}[!ht]
%\vspace{-10pt}
    \begin{center}
        \begin{tabular}{l}
            \includegraphics[scale = .45]{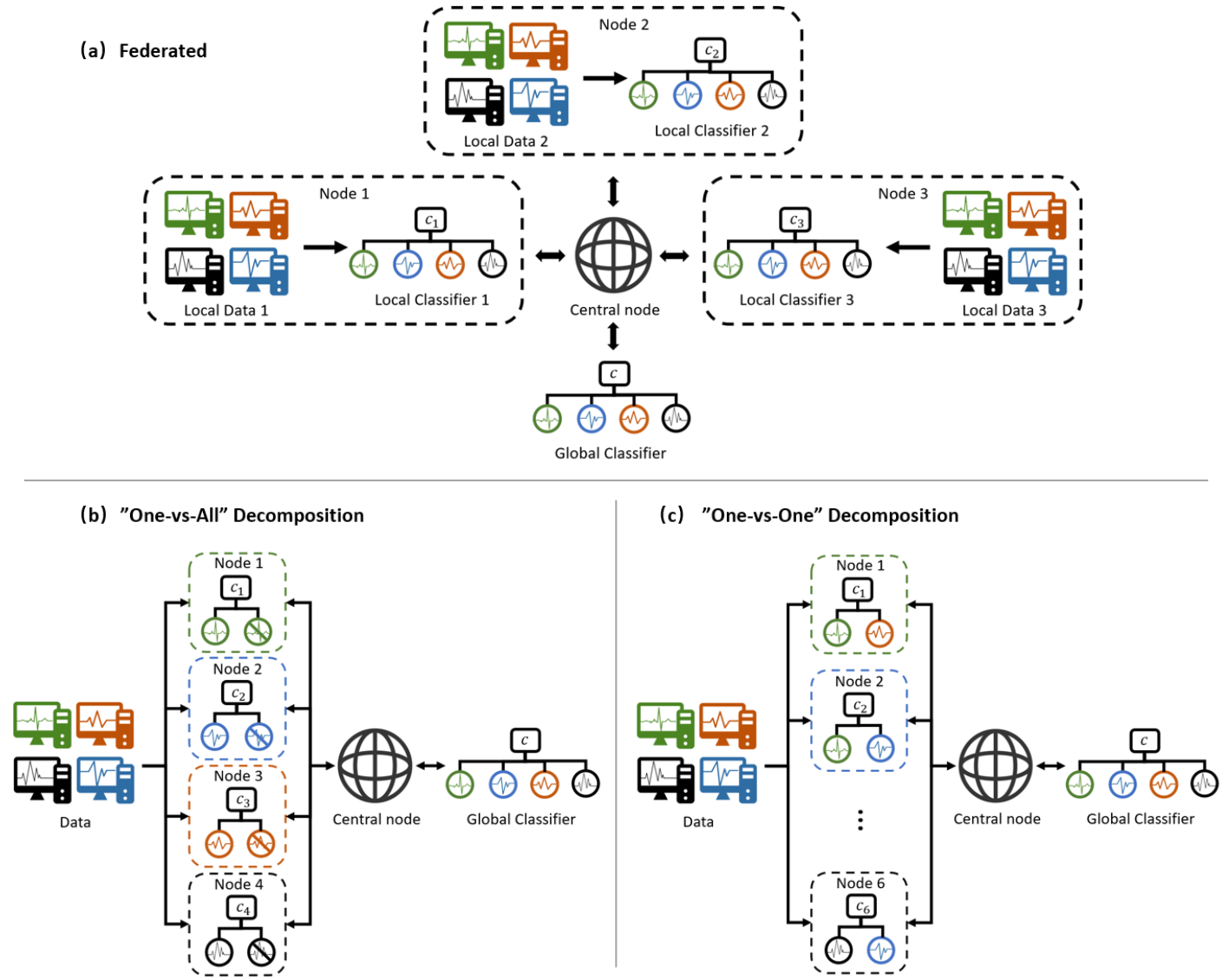}
        \end{tabular}
         \vspace{-5pt}
        \caption{Illustrations of existing multi-node multi-class classification approaches. Data in each local node only contains one type of attack. Based on the local data, only the binary classifiers can be obtained, denoted as $c_1, c_2$ and $c_3$. Our goal is to obtain the global classifier $c$.}\label{fig:existing}
    \end{center}
     \vspace{-10pt}
\end{figure}

\begin{figure}[!ht]
%\vspace{-10pt}
    \begin{center}
        \begin{tabular}{l}
            \includegraphics[scale = .35]{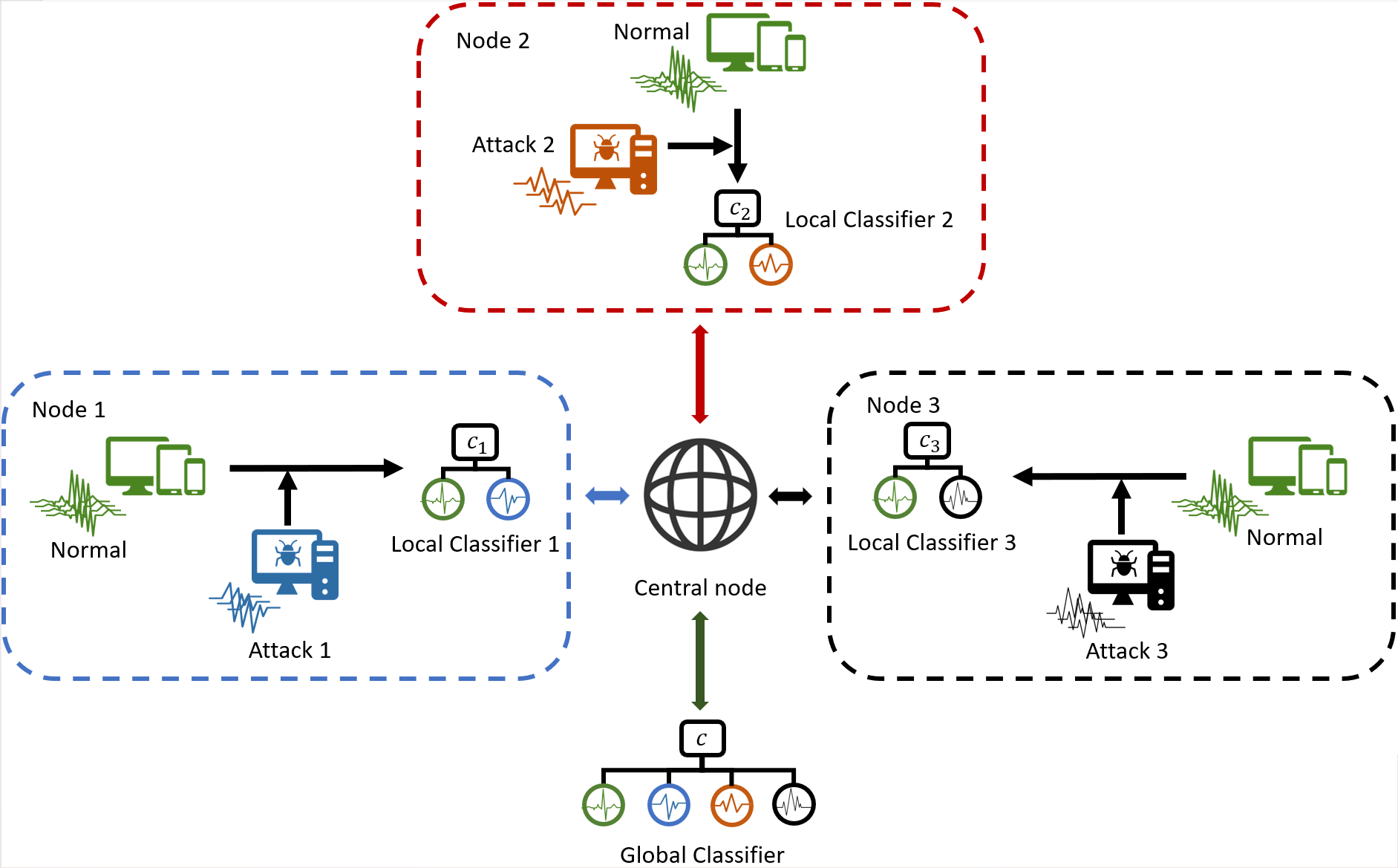}
        \end{tabular}
         \vspace{-5pt}
        \caption{An illustration of the attack detection setting with censoring nodes. Data in each local node only contains one type of attack. Based on the local data, only the binary classifiers can be obtained, denoted as $c_1, c_2$ and $c_3$. Our goal is to obtain the global classifier $c$.}\label{fig:setting}
    \end{center}
     \vspace{-10pt}
\end{figure}

Although extensive approaches have been developed for federated or decomposition multi-class classification in the past few years, most of the approaches require either of the following assumptions. (1) Data in each local node contains complete information for the multi-class classification, i.e., each node contains all classes of data so that a multi-class classifier can be calculated within each local node \cite{feng2020multi,roth2020federated}, and then the global classifier can be obtained through a federated average. (2) Data are shared across all local nodes so that the multi-class classifier can be obtained by either the one-vs-all decomposition scheme or the one-vs-one decomposition scheme, i.e., the multi-class classifier is obtained through an ensemble of multiple binary classifiers \cite{tax2002using,wu2003probability,galar2011overview}. Figure~\ref{fig:existing} illustrates the two typical types of existing approaches. Panel (a) in figure~\ref{fig:existing} shows the case in a federated approach, where assumption (1) is satisfied, and the local nodes can only access their own data. Panel (b) and (c) in figure~\ref{fig:existing} show the cases for an one-vs-all and an one-vs-one decomposition approach, respectively. In these cases, assumption (2) is satisfied.
However, such assumptions may not hold in real-world applications. For example, in disease screening studies, data collected from a basic medical center may only contain records from normal people and mild patients. On the other hand, in specialist hospitals, the majority of the records are from severe patients and critically ill patients. Such cases are also popular in intrusion-detection studies in IoT. In such studies, different attackers may affect different parts of the network, and carry out different types of attacks. As a result, within a certain node, there may only exist one typical type of attack \cite{chaabouni2019network,hajiheidari2019intrusion}. We now consider a toy example illustrated in figure~\ref{fig:setting}. Suppose there are three local nodes connected to a central node. Within each node, there exists normal data, shown as the green signals. There also exist three types of attackers, each of which is specified for a different local node. The attacked data is shown as the blue, red, and black signals, respectively. 
In such cases, information in each local node is incomplete, we thus call the local nodes ``censoring nodes''. Because of the data censoring, each local node can only train classifiers for its existing classes based on its data, such as the three local classifiers shown in figure~\ref{fig:setting}. However, in real-world applications, no one can guarantee that data from the missing classes will never appear in the future. For such data, the pre-trained classification model fails. The information for the missing classes needs to be obtained to compensate for the classification failure for local nodes and to obtain a global/complete classifier.  
One naive approach is pooling data from different nodes together to obtain the global multi-class classifier for all classes. However, in practice, data privacy and communication cost are crucial concerns. Sharing data across all nodes is infeasible \cite{rieke2020future, Li2020review}. There is an urgent need to learn the missing knowledge from other nodes without accessing their raw data. On the other hand, the data censoring is related to the practical network structures and node features, it is unlikely to obtain information on all possible combinations of pairs of classes. Therefore, the existing ensembling strategies described in \cite{tax2002using,wu2003probability,galar2011overview} are also infeasible.

\textbf{Our contributions.}
To combat the obstacles, we develop a novel multi-node multi-class classification ensemble approach with a combination of local classification with density estimation for multi-node attack detection, especially when data is censored within each local node. 
The approach is inspired by the ensemble learning approaches which integrate weak learners obtained in each local node to build a strong learner. Similarly, in the proposed multi-node multi-class classification ensemble approach, the ``weak learners'' can be regarded as the binary classifiers trained within each node using the local data; and the ``strong learner'' is the final multi-class classifier. Unlike traditional ensemble learning approaches, in which the local tasks within each node are based on full data, in the proposed approach, the local tasks are some ``partial'' tasks of the global multi-class classification task based on the censored local data. By combining local classification and density estimation, we obtain the informative parameter estimations for each class. We then complete the missing information of unobserved classes for each local node through communication with such estimations. The completed missing information can help each local node defend against potential attacks in the future, even if such types of attacks have not appeared in the past.
In addition, we show the proposed method is not specific to the attack detection problems. It is potentially suitable for general multi-class classification problems under data censoring with different types of data structures. In general, we summarize the contribution of the paper from three aspects.
%\vspace{-10pt}
\begin{enumerate}[noitemsep] 
    \item We propose a novel multi-node multi-class classification problem setting with censoring local nodes.
    \item We discuss several data/class censoring structures that are common in real-world applications.
    \item We propose an ensemble approach to aggregate local binary classifiers into a global multi-class classifier that is suitable for censoring nodes without sharing local data.
    \item The one-shot ensemble and the flexibility in the assumption of data structure make it possible to extend the proposed approach to a general multi-node multi-class classification problem under a federated learning framework.
\end{enumerate}
%\vspace{-10pt}

\section{Problem and Methodology}

\subsection{Traditional single-node and multi-node classification}
Suppose $J$ different nodes, denoted as $\mathrm{Node}_1, \ldots, \mathrm{Node}_J$, are connected to a central node $\mathrm{Node}_c$. Data $\{\x_{ij}, y_{ij}\}_{i=1}^{n_j}$ are collected from $\mathrm{Node}_j$, where $\x_{ij}\in \RR^d$ denotes the $d$-dimensional covariate, and $y_{ij}$ denotes the response label. We assume that there are $M+1$ classes in total, i.e., $y_{ij}\in\{0, 1, \ldots, M\}$. For example, in medical research, $y=0$ always represents the control group, and $y = 1, \ldots, M$ represent $M$ case groups. Another example is the Intrusion Detection System study, where $y=0$ represents normal/un-attacked records and $y = 1, \ldots, M$ represent $M$ attacks. %We further denote the sample size in the $j$-th node as $n_j$. 
Let $\mathcal{M}_j\subset \{0, 1, \ldots, M\}$ be the subset of classes exist in $\mathrm{Node}_j$. When we consider the classification problem within a single $\mathrm{Node}_j$, the local classifier $c_j^{(m_j)}(\x), m_j\in\mathcal{M}_j$ can be obtained through minimizing the following loss function,
\begin{equation}\label{eq:single}
c_j^{(m_j)}, m_j\in\mathcal{M}_j = \arg\min_{c_j^{(m_j)}, , m_j\in\mathcal{M}_j}\left\{-\sum_{i=1}^{n_j}\sum_{m_j\in\mathcal{M}_j}1(y_{ij}=m_j)\log c_j^{(m_j)}(\x_{ij})\right\} \quad \mathrm{s.t.} \quad \sum_{m_j\in\mathcal{M}_j} c_j^{(m_j)} = 1.
\end{equation}
For a traditional multi-node classification, the final goal is to obtain the $M+1$-class classifier $c^{(m)}(\x):=P(y=m|\x)$ for $m=0, 1, \ldots, M$, which minimizes the cross-entropy loss as follows,
\begin{equation}\label{eq:multiclass}
c^{(0)}, \ldots, c^{(m)} = \arg\min_{c^{(0)}, \ldots, c^{(m)}}\left\{-\sum_{j=1}^J\sum_{i=1}^{n_j}\sum_{m=0}^M1(y_{ij}=m)\log c^{(m)}(\x_{ij})\right\} \quad \mathrm{s.t.} \quad \sum_{m=0}^M c^{(m)} = 1,
\end{equation}
where the function $1(y_{ij}=m)$ is the indicator function that equals 1 when $y_{ij}=m$, otherwise 0.

\subsection{Ensemble multi-node multi-class classification}
Recall that in federated approaches, $\mathcal{M}_j = \{0, 1, \ldots, M\}$, thus the loss function for a global multi-class classification problem is simply a summation of all the loss functions of the local multi-class classification problems. It is easy to see that the global multi-class classifier can be obtained through a combination of all local $c_j$'s. On the other hand, in decomposition approaches, even though $c_j$'s within each local node are binary classifiers, the shared data and the complete information between one class to all other classes make it possible to expand the binary $c_j$'s to a global multi-class classifier.
However, we consider the cases when the local nodes are \textbf{censoring nodes}. Typically, we focus on the cases when each local node only contains data in two classes.
For example, if we follow the attack detection problem setting illustrated in figure \ref{fig:setting}, we assume within each local node, there exists un-attacked data ($y=0$) together with only one type of attack data ($y=m, m\in\{1, \ldots, M\}$).
According to the proposed data-censoring setting, $c_j$'s within each local node are binary one-vs-one classifiers. The access to the information of the unobserved classes for each local node is prevented by its censored data and the lack of enough pair-wise binary classifiers. Without the complete pair-wise information for all possible combinations of classes, it is challenging to obtain a feasible and effective global multi-class classifier based on the aggregation of limited local classifiers, especially when local data are not allowed to be shared. We thus need to extract more features from the local data, other than the estimated parameters for local classifiers, to communicate across all nodes so that the local nodes can complete their missing information. %we assume that there are only two classes within each single node. We consider a binary classification model for these two classes. For simplicity, we assume the response labels for these two classes are $0$ and $1$, respectively. In binary classification models, we aim to estimate the probability of $y=1$ through covariate $\x$. Consider the parametric classification model $P(y=1|\x) = f(\x; \beta)$ with unknown parameter $\beta$, if we take $f(\x; \beta) = \frac{\exp\{\x^\top\beta\}}{1+\exp\{\x^\top \beta\}}$, such a model then becomes the popular logistic regression model \cite{kleinbaum2002logistic}.

It is obvious that the global multi-class classifier $c^{(0)}, \ldots, c^{(m)}$ for classes $0, 1, \ldots, M$ depends on the information of the following two aspects: (1) the information of the response label $y$ for all classes, i.e. $P(y=m)$ for $m = 0, 1, \ldots, M$, and (2) the information of the predictor $\x$ for all classes. In each local node, we lack both of the aforementioned information for the missing classes. We thus need to borrow information from other nodes to fill in the missing parts. 
Intuitively, the information of the response label $P(y=m)$ can be provided from the local classifiers $c_j^{(m)}$'s, and the information of $\x$ is obtained by integrating the local density estimation of ${\x_{ij}}$ in each node. Specifically, the global classifier $c^{(m)}$ can be regarded as a summation of conditional probabilities $P(y=m, \mathrm{data}\in \mathrm{Node}_j |\x)$ for $j = 1, \ldots, J$. In addition, when the probability $P(\mathrm{data}\in \mathrm{Node}_j)$ and the distribution of $\x$ in $j$-th node are given, the conditional probability $P(y=m, \mathrm{data}\in \mathrm{Node}_j |\x)$ is proportional to $P(y=m| \mathrm{data}\in \mathrm{Node}_j, \x)$, which is the local classifier $c_j^{(m)}$. %We thus can fill in the first part of the missing information through the estimated parameters of the local binary classifiers. The distribution of $\x$ in $j$-th node can be obtained by density estimation. Therefore, the second part of missing information can be filled in through the parameters of the estimated local densities. In addition, the probability $P(\mathrm{data}\in \mathrm{Node}_j)$ is always regarded as the prior information of each node that is assumed to be known, or proportional to the local sample size. As a result, the ``joint information'' for the global multi-class classifier can be obtained by gathering the ``partial information'' provided by the local binary classifier and local density estimation from each local node. 
Motivated by this intuition, letting $f_{\x}^{(j)}$ be the distribution of $\x$ in $j$-th node, and $p_N^{(j)} = P(\mathrm{data}\in \mathrm{Node}_j), j = 1, 2, \ldots, J$, we derive the ensembling of local classifiers as
\[
c^{(m)} = \sum_{j=1}^Jw_{j,m}c_j^{(m)},
\]
where the weight $w_{j,m}\propto 1(m\in\mathcal{M}_j)g\left(f_{\x}^{(j)},p_N^{(j)}\right)$ with $1(\cdot)$ denoting the indicating function, and $g(\cdot,\cdot)$ denoting some unknown function. One possible choice of function $g$ is $g\left(f_{\x}^{(j)},p_N^{(j)}\right)=f_{\x}^{(j)}p_N^{(j)}$. We then state our main result as follows. 

\noindent\textbf{Main result} \textit{Suppose the local classifier for class $y=m$ obtained by minimizing \eqref{eq:single} in node $j$ is $\hat{c}_j^{(m)}$. Define $\mathcal{M}_j, f_{\x}^{(j)}$ and $p_N^{(j)}$ as above. Let $w_j = \frac{f_{\x}^{(j)}p_N^{(j)}}{\sum_{j=1}^Jf_{\x}^{(j)}p_N^{(j)}}$. If $w_j\ge \frac{1}{J}\left(\hat{c}_j^{(m)}\right)^{J-1}$, then we have that the following integrated classifier
\begin{equation}\label{eq:globalC}
    \hat{c}^{(m)} = \sum_{j=1}^J 1(m\in\mathcal{M}_j)w_j\hat{c}_j^{(m)}
\end{equation}
is a feasible solution of \eqref{eq:multiclass}.
}
The result is proved in Appendix \ref{app:proof}.

\subsection{Gaussian mixture model for single-node density estimation}
In practice, $p_N^{(j)}$ can be estimated using $\hat p_N^{(j)} = \frac{n_j}{\sum_j n_j}$, where $n_j$ is the sample size in $j$-th node. The remaining problem is then the density estimation for $f_{\x}^{(j)}$. In this paper, we use the Gaussian mixture model (GMM) \cite{reynolds2009gaussian} to obtain $\hat f_{\x}^{(j)}$. Specifically, we assume $\x$ in node $j$ follows a mixture of $K$ Gaussian components, i.e.
\begin{equation}\label{eq:GMM}
    f_{\x}^{(j)}(\x) = \sum_{k=1}^K\pi_k^{(j)}f_{\x|z_j=k}(\x),
\end{equation}
where $\x|z_j=k\sim N(\bm{\mu}_k^{(j)}, \Sigma_k^{(j)})$ comes from the $k$-th Gaussian component, and $z_j$ is the latent variable in node $j$ with $z_j = k$ indicating that the data comes from the $k$-th Gaussian component, and $\pi_k^{(j)} = P(z_j=k)$. Given the number of components $K$, model~\eqref{eq:GMM} can be estimated through the expectation maximization (EM) algorithm \cite{reynolds2009gaussian,biernacki2003choosing}. In GMM, a typical problem is choosing a proper number of components $K$ \cite{biernacki2003choosing}. In practice, we use a scree-plot of the likelihood score to determine the proper number of components.

\subsection{Main algorithm} 
Following the strategies above, the Ensemble Multi-node Multi-class Classification (EMMC) algorithm is detailed in algorithm~\ref{algo}.

\begin{algorithm}
\caption{Ensemble Multi-node Multi-class Classification Algorithm}
\label{algo}
\begin{algorithmic}
\State {\bfseries Input:} data collected from $J$ nodes, $\{\x_{ij}, y_{ij}\}_{i=1}^{n_j}, j = 1, \ldots, J$. 
\For{$j$ in $1,\ldots,J$}
   \State \textbf{Within $j$-th node:}
   \State \textbf{Step 1:} Fit a binary classification model based on $\{\x_{ij}, y_{ij}\}_{i=1}^{n_j}$. and obtain $\hat c_j^{(m)}(\x)$.
   \State \textbf{Step 2:} Fit a GMM for $\x_{ij}$, and obtain $\hat f_{\x}^{(j)}$ using EM algorithm.
   \State \textbf{Step 3:} Send $\hat c_j^{(m)}(\x)$, $\hat f_{\x}^{(j)}$, and $\hat p_N^{(j)} = \frac{n_j}{\sum_j n_j}$ to the central node. 
\EndFor
   \State \textbf{Within the central node:}
   \State \textbf{Step 4:} Calculate the global classifier $\hat{c}^{(m)}(\x)$ as in \eqref{eq:globalC}, for $m = 0, 1, \ldots, M$. 
   \State \textbf{Output:} $\hat{c}^{(m)}(\x), m = 0, 1, \ldots, M$.
\end{algorithmic}
\end{algorithm}

\section{Illustration of concept}
In this section, we use some examples to illustrate the proposed problem setting and approach. Particularly, we consider three typical structures for censoring local nodes. We then apply the proposed ensemble approach to both simulated examples and real-world datasets under each structure. The experiments were
conducted in Python using a Windows computer with 16GB of memory and an 8 core CPU.

\subsection{Simulated illustration}
In this section, we use a simulated toy example to illustrate the proposed approach. Consider data with four classes ($M=3$). Suppose the data for Class $m$ comes from a mixture distribution of $K_m$ Gaussian components with mean $\bm{\mu}_{k,m}$ and covariance matrix $\Sigma_{k,m}$, and each component is mixed with weight $\pi_{k,m}$. The specific parameter settings are listed in table~\ref{tab:parameter}.
\begin{table}[!ht]
\caption{Parameter settings for each class in the simulated illustration.}
\centering
\begin{tabular}{c|c|c|c|c}
\hline
class & Component & $\bm{\mu}_{k,m}$ & $\Sigma_{k,m}$ & $\pi_{k,m}$ \\ \hline
\multirow{2}{*}{class 0}  & 1  & $(-6,-1)^\top$ & \multirow{10}{*}{$I_2$} & 1/2 \\ \cline{2-3} \cline{5-5}  & 2  & $(-8,2)^\top$ &  & 1/2 \\ \cline{1-3} \cline{5-5} 
\multirow{3}{*}{class 1}                       & 1            & $(3,6)^\top$   &                             & 4/15   \\ \cline{2-3} \cline{5-5} 
                                               & 2             & $(4,4)^\top$   &                             & 6/15   \\ \cline{2-3} \cline{5-5} 
                                               & 3            & $(5,6)^\top$   &                             & 5/15   \\ \cline{1-3} \cline{5-5} 
\multirow{2}{*}{class 2}                       & 1             & $(0,-4)^\top$  &                             & 9/13   \\ \cline{2-3} \cline{5-5} 
                                               & 2             & $(0,-2)^\top$  &                             & 4/13   \\ \cline{1-3} \cline{5-5} 
\multicolumn{1}{c|}{\multirow{3}{*}{class 3}} & 1             & $(-1,5)^\top$  &                             & 6/15   \\ \cline{2-3} \cline{5-5} 
\multicolumn{1}{c|}{}                         & 2             & $(-2,4)^\top$  &                             & 5/15   \\ \cline{2-3} \cline{5-5} 
\multicolumn{1}{c|}{}                         & 3            & $(-2,5)^\top$  &                             & 4/15   \\ \hline
\end{tabular}
\label{tab:parameter}
\end{table}

We show the effectiveness of the proposed method through examples under three typical settings of data structure, which are detailed as follows. 
\begin{enumerate}
    \item[I] \textbf{(The ``star'' structure)} Consider there are three nodes ($J=3$). Data $\{\x, y\}$ within node $j$ comes from Class 0 and Class $j$, i.e., $y\in\{0, j\}, j=1,2,3$. 
    \item[II] \textbf{(The ``ring'' structure)} Consider there are four nodes ($J=4$). Data $\{\x, y\}$ within node $j$ comes from Class $j-1$ and Class $j$, i.e., $y\in\{j-1, j\}, j=1,\ldots,4$, where we define ``$y=4$'' is equivalent to ``$y=0$''. 
    \item[III] \textbf{(The ``fully-connected'' structure)} Consider there are six nodes ($J=6$). Within node $j$, we generate data $\{\x, y\}$ with $y\in\{j_1, j_2\}, j = 1, \ldots, 6$, where $\{j_1, j_2\}$ is one of all the possible combinations of 0, 1, 2, and 3, with $j_1\neq j_2$.
\end{enumerate}

\begin{figure}[!ht]
%\vspace{-10pt}
    \begin{center}
        \begin{tabular}{l}
            \includegraphics[scale = .37]{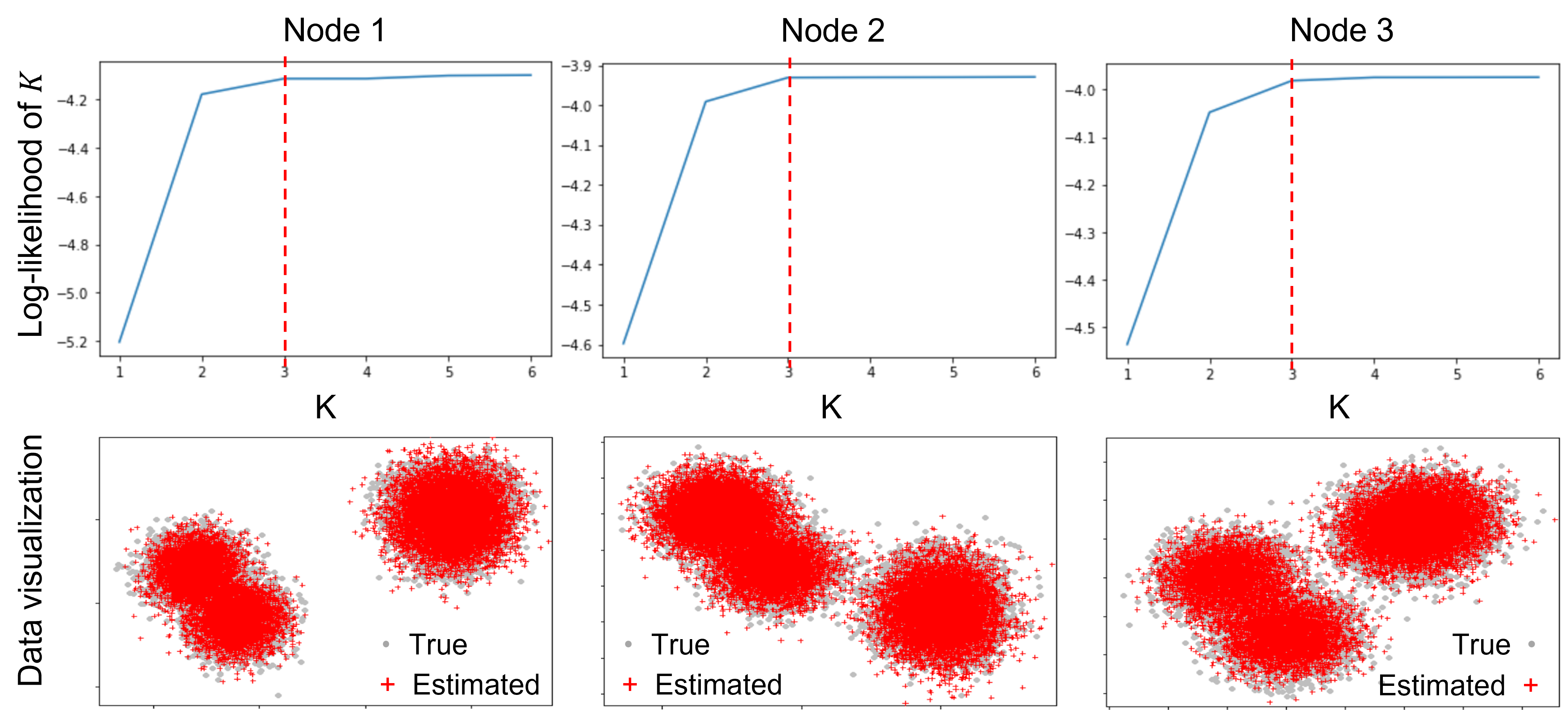}\\
        \end{tabular}
         \vspace{-5pt}
        \caption{Screeplot of the likelihood score for each node in the first setting.}\label{fig:screeplot}
    \end{center}
     \vspace{-10pt}
\end{figure}

We applied the proposed method to the simulated training data. For the first setting, we generated 14500, 13500, and 14500 samples within each node. Figure~\ref{fig:screeplot} shows the result of density estimation within each node. In figure~\ref{fig:screeplot}, the left, middle, and right columns each represent one single node respectively. The upper row shows the scree-plots of the likelihood scores. Based on the scree-plots, we chose the number of Gaussian components within each node to be $K = 3$. The bottom row of figure~\ref{fig:screeplot} visualizes the ground truth and estimated densities. We see that the generated samples from the estimated density (shown as the red ``+''s) perfectly match the samples generated from the true density (shown as the gray dots). Such an observation indicates the effectiveness of the density estimation through GMM. Similarly, based on the scree-plots of the likelihood scores, we chose $K = 3, 2, 2, 3$ for each node in the second setting, respectively, and $K = 3, 2, 2, 3, 3, 3$ for each node in the third setting, respectively.  %The results of density estimation for the second and third settings are shown in the Supplementary Materials.

After the density estimation through GMM, we applied the logistic regression model to the local data to build the binary classifier for each local node. Finally, we aggregated the estimated parameters of all local binary classifiers and the local densities to construct the global multi-class classifier as in EMMC algorithm~\ref{algo}. 

We generated 100, 150, 100, and 150 samples from Class 0, 1, 2, and 3, respectively, as the test data. We then carried out the classification tasks for the testing data under each of the settings. Specifically, we calculated both the precision and recall values defined as follows: 
$$ \mathrm{Precision} = \frac{\mathrm{True Positive}}{\mathrm{True Positive}+\mathrm{False Positive}},$$
and
$$ \mathrm{Recall} = \frac{\mathrm{True Positive}}{\mathrm{True Positive}+\mathrm{False Negative}}.$$
The results are listed in table~\ref{tab:simu}. The values outside the brackets are the mean precision and recall based on 50 replicates, and the values within the brackets are the corresponding standard deviations. From table~\ref{tab:simu}, we observe that under all three settings of data structure, the proposed method can provide a proper multi-class classifier with a high accuracy based on the binary classifiers provided by each local node.

\begin{table}[!ht]
\caption{Mean testing accuracy for each class under three settings with standard deviation in the brackets.}
\centering
\resizebox{0.5\textwidth}{!}{%
\begin{tabular}{r|llll}
\hline
              & Class 0  & Class 1 & Class 2 & Class 3 \\ \hline
              & \multicolumn{4}{c}{Precision}\\
I   & $99.9\%(.29\%)$ & $98.62\% (.84\%)$   &  $100\%(0\%)$           &  $99.13\%(.72\%)$         \\ %\hline
II   &   $99.8\%(.39\%)$         & $99.27\%(.46\%)$         &    $100\%(0\%)$         &  $98.68\%(.92\%)$          \\ %\hline
III &   $99.9\%(.29\%)$         & $99.53\% (.42\%)$          &   $100\%(0\%)$          &    $99.2\%(.49\%)$        \\ \hline
              & \multicolumn{4}{c}{Recall}\\
I   & $100\%(0\%)$ & $99.2\% (.65\%)$   &  $99.8\%(.40\%)$           &  $98.6\%(.87\%)$         \\ %\hline
II   &   $100\%(0\%)$         & $98.67\%(.94\%)$         &    $99.9\%(.30\%)$         &  $99.2\%(.50\%)$          \\ %\hline
III &   $100\%(0\%)$         & $99.2\% (.50\%)$          &   $100\%(0\%)$          &    $99.47\%(.50\%)$        \\ \hline
\end{tabular}%
}
\label{tab:simu}
\end{table}

In practice, the three types of structures may suitable for different scenarios. In the cases of attack detection, it is obvious that the class $y = 0$, i.e., the normal/un-attacked data exists within all local nodes. Thus the data structure is by nature the ``star'' structure. In the following section, the proposed approach was applied to two real-world attack-detection datasets for further illustration.

\subsection{An example for the NUSW-NB15 dataset}
UNSW-NB15 is a dataset for cyber security research provided by \cite{Moustafa2015}. The dataset contains normal/un-attacked data and data of nine types of attacks: Fuzzers, Analysis, Backdoors, DoS, Exploits, Generic, Reconnaissance, Shellcode, and Worms. Denote the nine types of attacks as $y = 1, \ldots, 9$, respectively. We applied the proposed multi-node multi-class classification ensemble approach to see if the nine classes with $y = 1, \ldots, 9$ and the normal data ($y=0$) can be identified. Besides the training data, UNSW-NB15 also provides a test dataset. Recall that the proposed approach is based on the estimation of the data distribution, so a basic assumption is that there exists no distribution drift between the training and test data. However, in practice, whether there exists a distribution drift between the training data and the testing data is unknown. We thus need to consider a hypothesis testing for the distribution drift first. Denote the distribution of the training data and the testing data as $p_{\x}^{Tr}$ and $p_{\x}^{Te}$, respectively. The difference between the distributions $p_{\x}^{Tr}$ and $p_{\x}^{Te}$ can be measured by the Kullback–Leibler (KL) divergence $KL(p_{\x}^{Tr}\|p_{\x}^{Te})$ \cite{kullback1951information}. It is known that if $p_{\x}^{Tr} = p_{\x}^{Te}$, the KL divergence $KL(p_{\x}^{Tr}\|p_{\x}^{Te}) = 0$. We considered the following hypothesis testing,
\[
H_0: KL(p_{\x}^{Tr}\|p_{\x}^{Te}) = 0; \quad H_a:KL(p_{\x}^{Tr}\|p_{\x}^{Te}) > 0.
\]
Notice $KL(p_{\x}^{Tr}\|p_{\x}^{Te})$ is calculated based on the true distributions, which are unobserved. We can only obtain the estimation of $KL(p_{\x}^{Tr}\|p_{\x}^{Te})$, denoted as $\widehat{KL}$ from the observed samples. Moreover, the distribution of $\widehat{KL}$ under the null hypothesis (the ``null distribution'') can be complicated and difficult to parameterize. We thus applied the non-parametric testing by obtaining the empirical null distribution of $\widehat{KL}$ through randomly splitting the pooled dataset of the training and test data repeatedly, and calculating the $\widehat{KL}$s between the split datasets. We repeated the splitting 100 times and calculated the proportion of the $\widehat{KL}$s being greater than the original one as the testing p-value. We also carried out the hypothesis testing on each class separately, and the results are listed in table~\ref{tab:drifttesting}. From the results, we can see that there is a significant distribution drift between the training and test data. Specifically, the differences is significant in the normal data and the data of attack Fuzzers, Analysis, Backdoor, and Generic. Therefore, we only use the given training dataset in the experiment.

\begin{table}[H]
\centering
\caption{P-values for the hypothesis testings of the distribution drifts between the training and testing datasets.}
\resizebox{\textwidth}{!}{%
\begin{tabular}{l|lllllllllll} \hline
Class   & \textbf{Overall} & Normal & Fuzzers & Analysis & Backdoor & DoS & Exploits  & Generic & Reconnaissance & Shellcode & Worms \\ \hline
P-value & \textbf{0.02} & 0.02 & 0.00 & 0.00 & 0.00 & 0.36 & 0.48 & 0.00 & 0.30 & 0.62 & 0.26
  \\ \hline
\end{tabular}%
}
\label{tab:drifttesting}
\end{table}

Since the proposed approach is based on an ensemble of the local classifiers, we need to check the sample size of each type of attack to see if it is enough to obtain a good local classifier. The sample size for each type of attack is listed in table~\ref{tab:samplesize}. We observe that for the attack ``Worms'', there are only 44 records, which is too small compared to the rest of attacks. Thus we abandoned it in the experiment. Moreover, data balancing is also important for training a good classifier. However, the dataset at hand highly skewed. Therefore, we balanced each type of attacks using the following strategy: if the sample size of the attack is less than 1000, we keep the data untouched; if the sample size is greater than 1000, we randomly pick out 1000 representatives. Note that NUSW-NB15 is a dataset for attack detection, so we considered the ``star'' structure, i.e. there are nine local nodes (J=9), and data $\{\x, y\}$ within node $j$ satisfies $y\in\{0, j\}, i=1, \ldots, 8$. Again, to ensure the balance between the normal and the attacked data within each local node, we sub-sampled the normal data according to the sample size of each type of attack, and assigned them to all the nodes. 

\begin{table}[H]
\centering
\caption{Sample sizes for each type of attacks.}
\resizebox{\textwidth}{!}{%
\begin{tabular}{l|lllllllll} \hline
Attack & Fuzzers & Analysis & Backdoors & DoS & Exploits & Generic & Reconnaissance & Shellcode & Worms \\ \hline
Samplesize & 6062 & 677 & 583 & 4089 & 11132 & 18871 & 3496 & 378 & 44
  \\ \hline
\end{tabular}%
}
\label{tab:samplesize}
\end{table}

\begin{figure}[!ht]
%\vspace{-10pt}
    \begin{center}
        \begin{tabular}{l}
            \includegraphics[scale = .45]{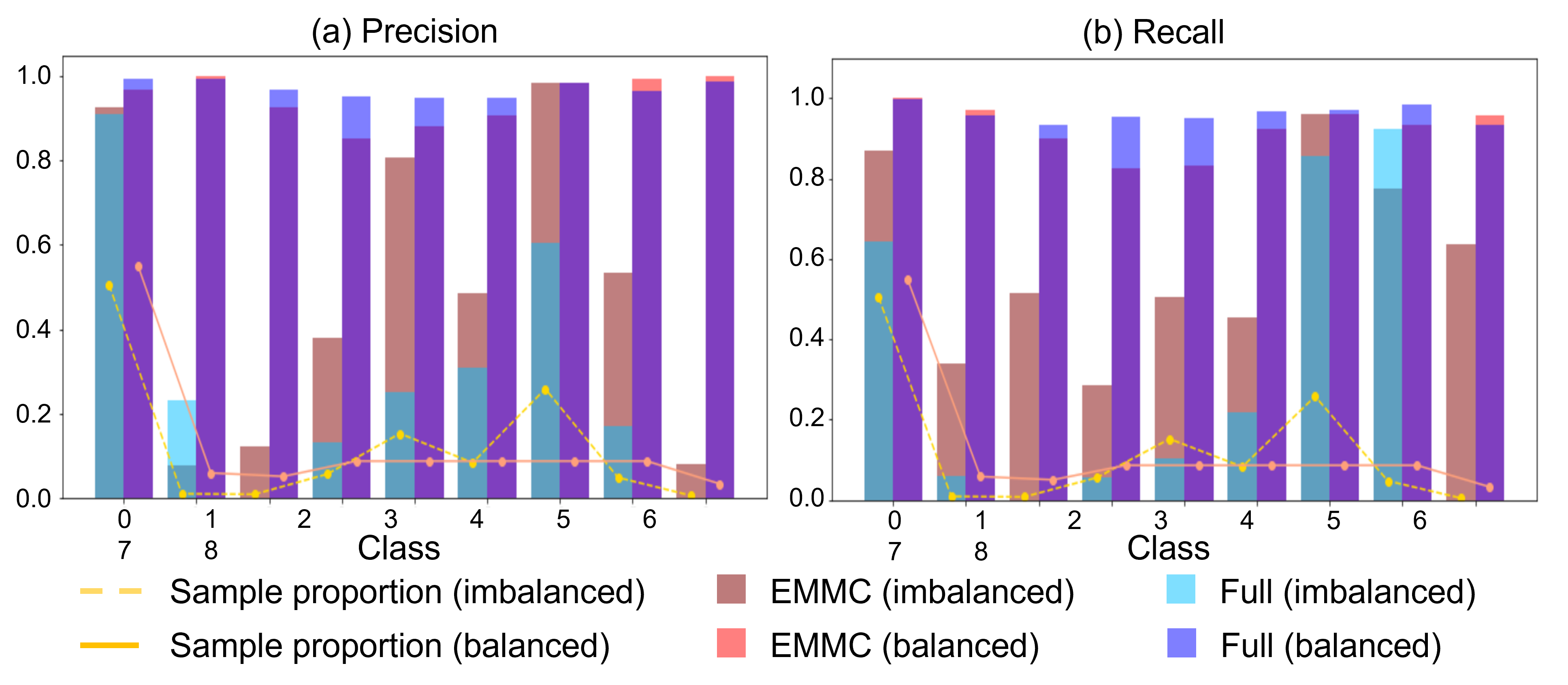}\\
        \end{tabular}
         \vspace{-5pt}
        \caption{Comparison of the testing precisions and recalls of the ensemble (EMMC)/full-data (Full) approaches for the imbalanced and balanced data.}\label{fig:compare}
    \end{center}
     \vspace{-10pt}
\end{figure}

We randomly picked $75\%$ of the pre-processed dataset to train the proposed classifier, and used the remaining $25\%$ as the testing set. We calculated the precision and recall of each class in the testing dataset for both the balanced data and the original imbalanced data (with ``Worms'' deleted) to see the effect of the data balancing. The comparison is shown in figure~\ref{fig:compare}. Panel (a) shows the precisions and panel (b) shows the recalls. The yellow lines are the proportion of each class in the training data, with the dashed lines representing the imbalanced data and the solid lines representing the balanced data. The sample size for Class 0 is a combination of all nodes, thus the proportion is greater than other classes. We observe that after balancing, the lines become much flatter. The groups of bars represent the test precisions/recalls of each class. The red/purple bars show the ensemble/full-data results for the balanced data, and the brown/blue bars show the ensemble/full-data results for the imbalanced data. We can observe that even though the sample sizes become smaller than the original ones for some classes after balancing, the results are much better than those of the imbalanced data. Such an observation verifies the necessity of data balancing. We then focus on the balanced data from now on.

Similar to the simulated example, we plot the scree-plots of the likelihood score, and chose $K = 35$ for the GMM in each local node. For the local binary classifiers, we applied four commonly used machine learning approaches: Logistic Regression (LR), Random Forest (RF), Support Vector Classifier (SVC), and Naive Bayes with Gaussian (NB). After 50 replicates, the testing precisions and recalls for each class are listed in table~\ref{tab:NUSW-NB15}. The results show that the proposed approach is feasible and effective when the training dataset is balanced.

\begin{table}[htp]
%\begin{center}
\caption{Mean classification precision and recall using different local classifiers with standard deviation of the balanced dataset}
\centering
\resizebox{\textwidth}{!}{%
    \begin{tabular}{r|lllllllll}
        \hline
        %\multicolumn{8}{c}{\textbf{Federated learning}}\\ \hline
          &  Normal & Analysis & Backdoor & DoS & Exploits &Fuzzers & Generic & Reconnaissance& Shellcode \\ \hline 
          & \multicolumn{8}{c}{Precision}\\
     LR & $99.98\%(.29\%)$ & $96.25\%(1.91\%)$ & $87.63\%(3.61\%)$ & $80.70\%(3.81\%)$ & $81.67\%(5.94\%)$& $88.55\%(2.34\%)$& $95.70\%(1.08\%)$& $91.56\%(6.29\%)$& $91.42\%(3.37\%)$  \\
         RF & $100\%(0\%)$ & $96.18\%(1.82\%)$ & $87.15\%(3.99\%)$ & $81.21\%(3.02\%)$ & $80.87\%(5.28\%)$& $88.31\%(4.14\%)$& $95.54\%(1.02\%)$& $92.77\%(5.58\%)$& $89.87\%(3.53\%)$  \\
         
         SVC & $99.60\%(.24\%)$ & $96.28\%(1.85\%)$& $86.84\%(4.20\%)$& $80.39\%(3.43\%)$& $78.98\%(7.08\%)$& $88.50\%(2.57\%)$& $95.38\%(1.05\%)$& $91.08\%(10.60\%)$& $89.63\%(4.9\%)$ \\
         NB & $100\%(0\%)$ & $96.08\%(1.96\%)$& $87.44\%(3.89\%)$& $81.07\%(4.23\%)$& $80.15\%(7.97\%)$& $86.91\%(5.29\%)$& $95.69\%(.91\%)$& $89.90\%(10.48\%)$& $91.22\%(3.50\%)$ \\         
     \hline
          \multicolumn{10}{c}{Recall}\\
         LR& $96.57\%(.37\%)$ & $98.78\%(1.31\%)$ & $95.40\%(1.90\%)$& $85.25\%(2.84\%)$ & $80.80\%(4.64\%)$ & $87.50\%(7.30\%)$& $98.59\%(1.80\%)$& $99.58\%(.79\%)$& $99.54\%(1.01\%)$ \\
         
         RF& $96.38\%(4.73\%)$  & $99.13\%(1.14\%)$ & $ 95.66\%(3.39\%)$ & $83.27\%(4.49\%)$& $81.43\%(5.78\%)$& $88.37\%(5.50\%)$& $99.01\%(1.07\%)$ & $99.62\%(.96\%)$ & $99.40\%(1.82\%)$ \\
         
         SVC& $96.69\%(.44\%)$& $97.86\%(1.90\%)$  & $95.21\%(3.33\%)$ & $82.98\%(4.57\%)$ & $77.44\%(4.98\%)$& $87.12\%(8.03\%)$& $98.37\%(1.50\%)$& $98.97\%(2.94\%)$ & $99.83\%(.54\%)$\\         
         
         NB& $96.77\%(.38\%)$& $99.48\%(.88\%)$  & $93.83\%(3.60\%)$ & $84.81\%(3.82\%)$ & $78.90\%(7.3\%)$& $85.69\%(9.32\%)$& $98.88\%(1.23\%)$& $98.80\%(2.24\%)$ & $100\%(0)$ \\         
         \hline
         
         \hline

        \end{tabular}
}
%\end{center}
\label{tab:NUSW-NB15}
\end{table}

\subsection{An example for The EdgeIIoT dataset}
In \cite{edge}, a new comprehensive and realistic cyber security dataset of IoT and IIoT, dubbed Edge-IIoT, is introduced. The dataset is constructed for machine-learning-based intrusion detection systems. Twelve attack scenarios, namely "Backdoor" , "DDoS-HTTP" , "DDoS-ICMP" , "DDoS-TCP" , "DDoS-UDP"  , "Password" , "Port-Scanning" ,"Ransomware", "SQL-injection" , "Uploading" , "Vulnerability-scanner" and "XSS", are covered in the dataset ($M=12$). We denote the response label as $y = 1, \ldots, 12$ for each attack, respectively. Similar to the previous example, we consider the ``star'' structure: in the $j$-th node, we consider a binary classification task for $y=0$ (the normal data) and $y=j$.

\begin{figure}[!ht]
%\vspace{-10pt}
    \begin{center}
        \begin{tabular}{l}
            \includegraphics[scale = .43]{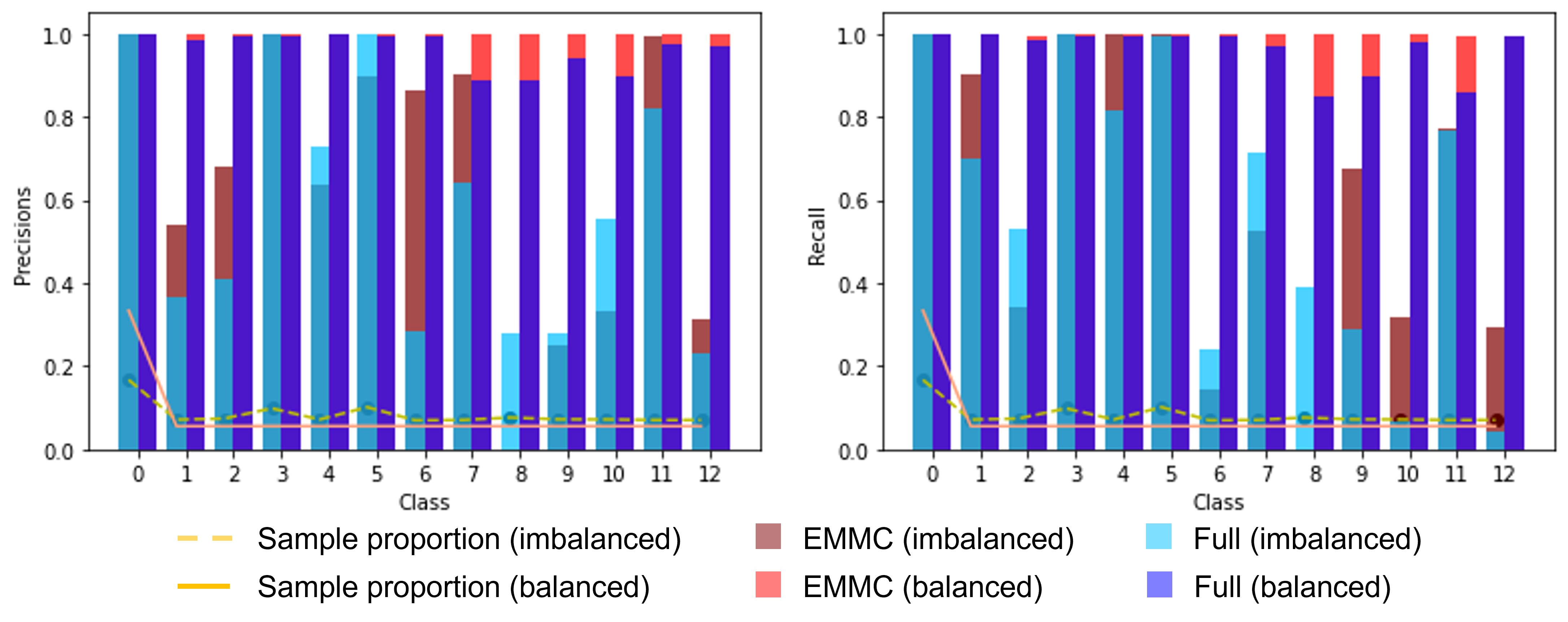}\\
        \end{tabular}
         \vspace{-5pt}
        \caption{Comparison of the testing precisions and recalls of the ensemble (EMMC)/full-data (Full) approaches for the imbalanced and balanced data.}\label{fig:compare2}
    \end{center}
     \vspace{-10pt}
\end{figure}

We randomly picked $75\%$ of the dataset to train the classifiers and used the remaining $25\%$ as the testing set. 
Again, we observe that the attacked data are imbalanced for different classes, such skewness lowers the accuracy of the local classifiers, as shown in figure~\ref{fig:compare2}. To balance the dataset, we randomly selected 1000 normal data and 1000 attacked data for each node. %These data are seen as the original data set. 
Based on the scree-plots of the log-likelihood scores, the number of Gaussian components in GMM were set to be $K=15$. For local binary classifiers, we applied the machine learning approaches support vector machine (SVM),  k-nearest neighbor (kNN), Random Forest (RF), and Logistic Regression (LR) as in \cite{edge}. Averaged over 50 replicates, the precisions and recalls for each class are listed in table~\ref{tab:Edgellot}. The results again support the feasibility and effectiveness of the proposed approach.

\begin{table}[htp]
%\begin{center}
\caption{Mean classification precision and recall using different local classifiers with standard deviation of the balanced dataset}
\centering
\resizebox{\textwidth}{!}{%
    \begin{tabular}{r|lllllllllllll}
        \hline
        %\multicolumn{13}{c}{\textbf{Ensembled Multi-class Classification}}\\ \hline
          &  Benign & Backdoor & DDoS-HTTP & DDoS-ICMP& DDoS-TCP & DDoS-UDP & Password & Port-Scanning & Ransomware & SQL-injection & Uploading & Vulnerability-scanner & XSS  \\ \hline 
          & \multicolumn{13}{c}{Precision}\\
         SVM & $99.89\%(.04\%)$ & $99.89\%(.21\%)$ & $100\%(0)$ & $100\%(0)$ & $100\%(0)$& $100\%(0)$& $100\%(0)$& $100\%(0)$& $99.98\%(.08\%)$ & $100\%(0)$& $100\%(0)$& $100\%(0)$& $100\%(0)$ \\
         kNN & $99.86\%(.09\%)$ & $99.77\%(.47\%)$ & $100\%(0)$ & $100\%(0)$ & $100\%(0)$& $99.79\%(.91\%)$& $100\%(0)$& $100\%(0)$& $99.97\%(.08\%)$ & $100\%(0)$& $100\%(0)$& $100\%(0)$& $100\%(0)$ \\
         RF & $99.89\%(.06\%)$ & $100\%(0)$& $100\%(0)$& $100\%(0)$& $100\%(0)$& $100\%(0)$& $100\%(0)$& $100\%(0)$& $100\%(0)$& $100\%(0)$& $100\%(0)$& $100\%(0)$& $100\%(0)$ \\
         LR & $99.89\%(.06\%)$ & $100\%(0)$& $100\%(0)$& $100\%(0)$& $100\%(0)$& $100\%(0)$& $100\%(0)$& $100\%(0)$& $100\%(0)$& $100\%(0)$& $100\%(0)$& $100\%(0)$& $100\%(0)$ \\         
     \hline
          \multicolumn{13}{c}{Recall}\\
         SVM& $99.99\%(.01\%)$  & $99.90\%(.27\%)$ & $ 99.76\%(.41\%)$ & $100\%(0)$& $100\%(0)$& $100\%(0)$& $99.83\%(.23\%)$ & $100\%(0)$ & $99.98\%(.08\%)$ & $99.93\%\%(.14\%)$ & $99.85\%(.22\%)$ & $99.68\%(.34\%)$& $99.80\%(.29\%)$\\
         
         kNN& $99.96\%(.11\%)$  & $99.88\%(.30\%)$ & $ 99.61\%(.47\%)$ & $100\%(0)$& $100\%(0)$& $100\%(0)$& $99.73\%(.32\%)$ & $100\%(0)$ & $100\%(0)$ & $99.93\%\%(.14\%)$ & $99.94\%(.14\%)$ & $99.56\%(.53\%)$& $99.76.31\%(.29\%)$\\
         
         RF& $100\%(0)$& $99.96\%(.11\%)$  & $99.76\%(.41\%)$ & $100\%(0)$ & $100\%(0)$& $100\%(0)$& $99.85\%(.23\%)$& $100\%(0)$ & $99.98\%(.08\%)$ & $99.93\%(.14\%)$ & $99.88\%\%(.22\%)$ & $99.56\%(.48\%)$ & $99.76\%(.29\%)$\\         
         
         LR& $100\%(0)$& $99.98\%(.08\%)$  & $99.74\%(.45\%)$ & $100\%(0)$ & $100\%(0)$& $100\%(0)$& $99.85\%(.23\%)$& $100\%(0)$ & $100\%(0)$ & $99.91\%(.16\%)$ & $99.89\%\%(.21\%)$ & $99.68\%(.34\%)$ & $99.62\%(.04\%)$\\         
         \hline
         
         \hline

        \end{tabular}
}
%\end{center}
\label{tab:Edgellot}
\end{table}

When dealing with general multi-class classification problems, such as classification for digital numbers (e.g. the MNIST dataset), we have no obvious justification on assuming a ``star'' structure any more. The other two structures (``ring'' and ``fully-connected''), or some structures (e.g., each node containing two classes randomly) are also possible. For simplicity, we focus on the ``star'' structure of real-world datasets for attack detection in the following sections. We refer to Appendix \ref{app:mnist} for general classification problems with the ``ring'' and ``fully-connected'' structures.

\section{An Analysis of the MQTT Protocol-based Dataset}

Message Queuing Telemetry Transport (MQTT) protocol is one of the most commonly used protocol in IoT \cite{singh2015secure,gupta2021mqtt}.
We consider the dataset provided by \cite{hindy2020mqtt}, which is the first simulated dataset under MQTT protocol \footnote{\href{https://ieee-dataport.org/open-access/mqtt-internet-things-intrusiondetection-dataset}{https://ieee-dataport.org/open-access/mqtt-internet-things-intrusiondetection-dataset}}. 
In this dataset, four different types of attacks are considered: Aggressive scan (Scan-A), User Datagram Protocol (UDP) scan (Scan-sU), Sparta SSH brute-force (Sparta), and MQTT brute-force attack (MQTT-BF). Therefore, we have $M=4$, with $y=1, 2, 3, 4$ indicating the Scan-A, Scan-sU, Sparta, and MQTT-BF attacks, respectively. 
Besides the multi-class dataset, datasets for each type of attack are also provided. We call such datasets the ``binary'' datasets. Each binary dataset contains records for the normal data and one certain type of attack. Moreover, among the simulated packet-based features, unidirectional-based features, and bidirectional-based features, we focus on the bidirectional-based features, which can provide the highest classification accuracy in previous related works in the experimental studies. 

\subsection{Feature subspace for each type of attacks}
We first try to understand the mechanism of each type of attack by extracting the ``feature subspace'' for each type of attack. Here we refer to the subspace spanned by certain linear combinations of original features as the ``feature subspace'', which differs most between the attacked and normal data.

\subsubsection{Feature subspace extracting}
To extract the feature subspace, we applied the principal optimal transport direction (POTD) approach \cite{meng2020sufficient}. We summarize the POTD approach as follows. Firstly, an optimal transport plan \cite{villani2009optimal,peyre2019computational,zhang2021review,zhang2022projection} is calculated between the attack data $\x|y = m, m = 1, 2, 3, 4$ and the normal data $\x|y=0$. Secondly, a ``displacement matrix'' $P_{m,0}, m=1,2,3,4$, which represents the empirical transport plan from $\x|y=m$ to $\x|y=0$, is used to measure the difference between the attack and normal data. POTD approach then extracts the major directions of $P_{m,0}$ using the principal component analysis (PCA). The major directions, denoted as $B_m, m=1,2,3,4$, are proved to be the directions that differ most between the attack and normal data. The feature subspace for each type of attack is defined as the subspace spanned by $B_m$. 

\begin{figure}[!ht]
%\vspace{-10pt}
    \begin{center}
        \begin{tabular}{l}
            \includegraphics[scale = .55]{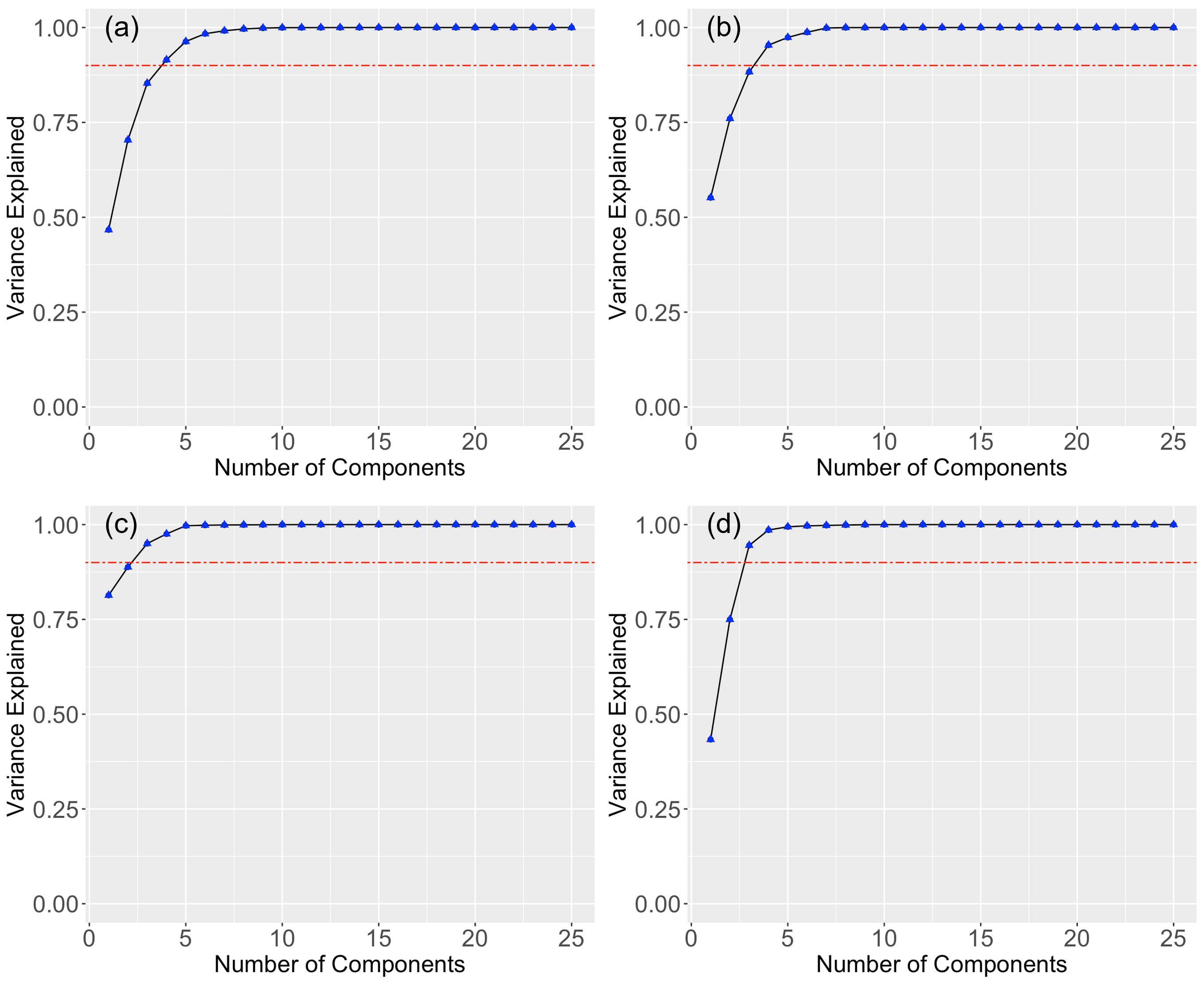}\\
        \end{tabular}
         \vspace{-5pt}
        \caption{Scree-plots for the PCA of the displacement matrix $P_{m,0}$. Panel (a): $m=1$; Panel (b): $m=2$; Panel (c): $m=3$; and Panel (d): $m=4$.}\label{fig:POTDscree}
    \end{center}
     \vspace{-10pt}
\end{figure}

\begin{figure}[!ht]
%\vspace{-10pt}
    \begin{center}
        \begin{tabular}{l}
            \includegraphics[scale = .3]{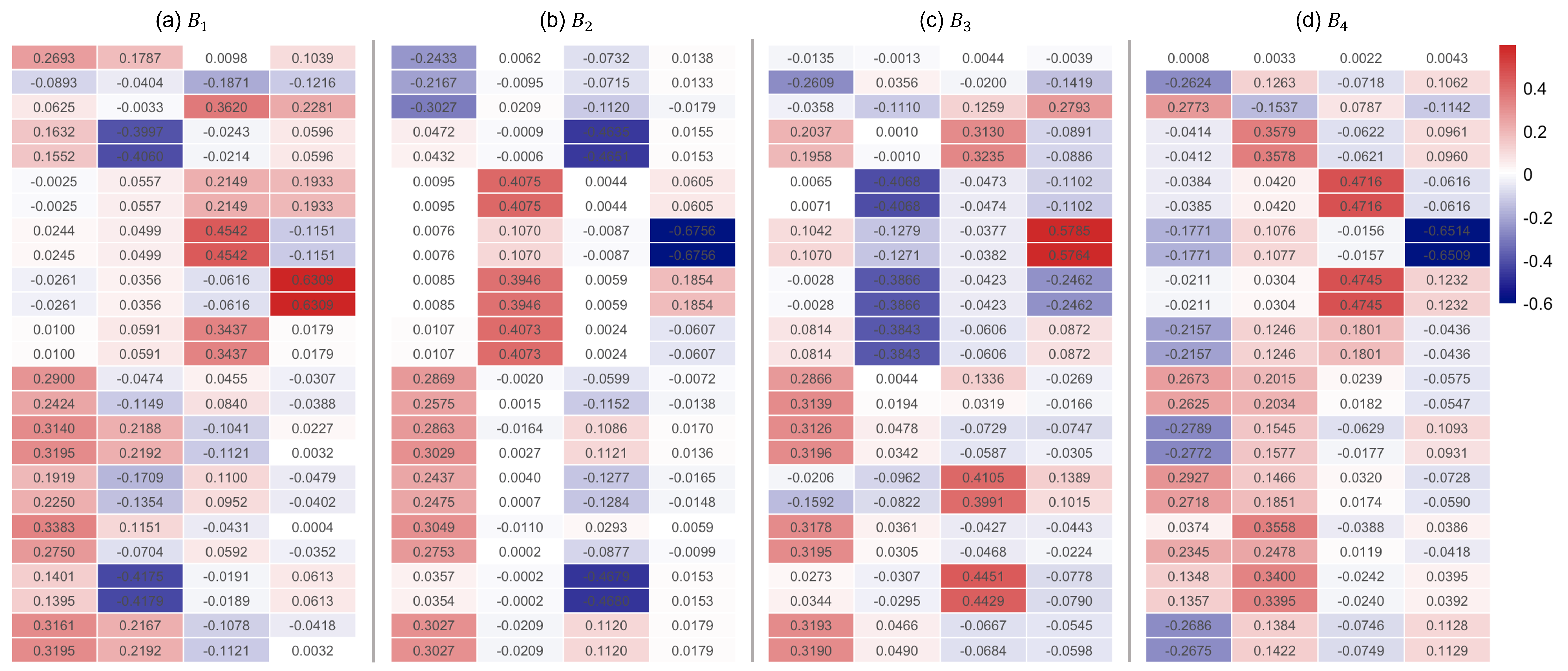}\\
        \end{tabular}
         \vspace{-5pt}
        \caption{Heatmaps of the four direction matrices.}\label{fig:POTDheatmap}
    \end{center}
     \vspace{-10pt}
\end{figure}

Based on the scree-plots for the PCA of $P_{m,0}$ shown in figure~\ref{fig:POTDscree}, the first four principal components can explain over $90\%$ of the information of the original matrix $P_{m,0}$. We thus set the dimension for direction matrix $B_m$ to be $d=4$, for all $m=1,\ldots, 4$. The direction matrices for the four types of attacks are shown in figure~\ref{fig:POTDheatmap}. From the figure, we observe that the feature subspaces are specified for different types of attacks. We further study the specificity of the feature subspaces using more statistical tools.

\subsubsection{Specificity of feature subspaces}

We show the specificity of the four extracted feature subspaces by comparing the distributions of the Wasserstein distances \cite{villani2009optimal,peyre2019computational,zhang2021review,zhang2022projection} between the attack and normal data after projecting data onto each feature subspace. Specifically, we obtain the empirical distributions as follows,
\begin{enumerate}
    \item Randomly sample 1000 attacked samples and 1000 normal samples from the four binary datasets.
    \item Project the random samples onto each feature subspace.
    \item Calculate the Wasserstein distance between the projected attack and normal samples.
    \item Repeat steps (1)-(3) for $100$ times.
\end{enumerate}

\begin{figure}[!ht]
%\vspace{-10pt}
    \begin{center}
        \begin{tabular}{l}
            \includegraphics[scale = .4]{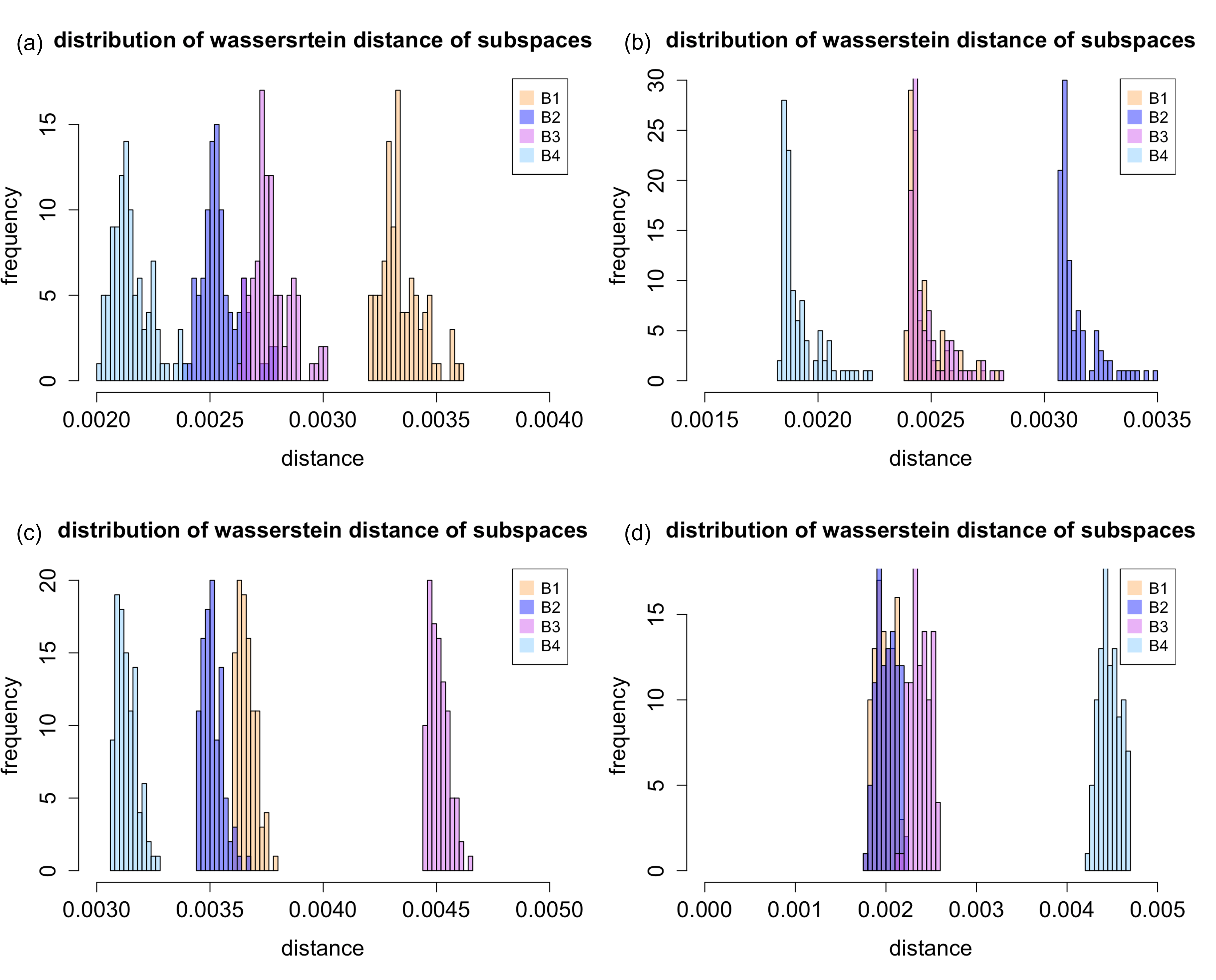}\\
        \end{tabular}
         \vspace{-5pt}
        \caption{Comparison of empitical distributions of the Wasserstein distance between the attacked data and normal data after projecting onto four subspaces. Panel (a): scan\_A; Panel (b): scan\_sU; Panel (c): Sparta; and Panel (d): MQTT BF.}\label{fig:WD_hist}
    \end{center}
     \vspace{-10pt}
\end{figure}

Intuitively, if the extracted feature subspaces are specified for the corresponding attack types, we expect that the empirical distribution of the Wasserstein distance between the attacked data and the normal data after projecting onto the corresponding feature subspace will be clearly separated from those projected onto all other subspaces. The histograms for the Wasserstein distances are shown in Figure~\ref{fig:WD_hist}. From the figure, we observe that it's clear that after projecting, the Wasserstein distances between the $m$-th attack and the normal data in the feature subspace spanned by $B_m$ are significantly larger than others $(m=1,2,3,4)$. Such an observation matches the aforementioned intuition. It verifies the specificity of the four extracted feature subspaces. 

\subsection{Attack detection}
The specificity shown above implies that the four types of attacks may be under different mechanisms. As a result, attack detection using a multi-class classification approach becomes more feasible. Similar to the data pre-processing in \cite{article1}, we excluded the IP addresses and protocol from the datasets. 
We considered each binary dataset as a node in the experimental studies. In other words, we considered the ``star'' structure as in the previous simulation studies and real-world data illustrations. In such a structure, there are four nodes (J = 4). Each node only contains information to identify one of the Scan-A, Scan-sU, Sparta, and MQTT-BF attacks. 

\begin{figure}[!ht]
%\vspace{-10pt}
    \begin{center}
        \begin{tabular}{l}
            \includegraphics[scale = .5]{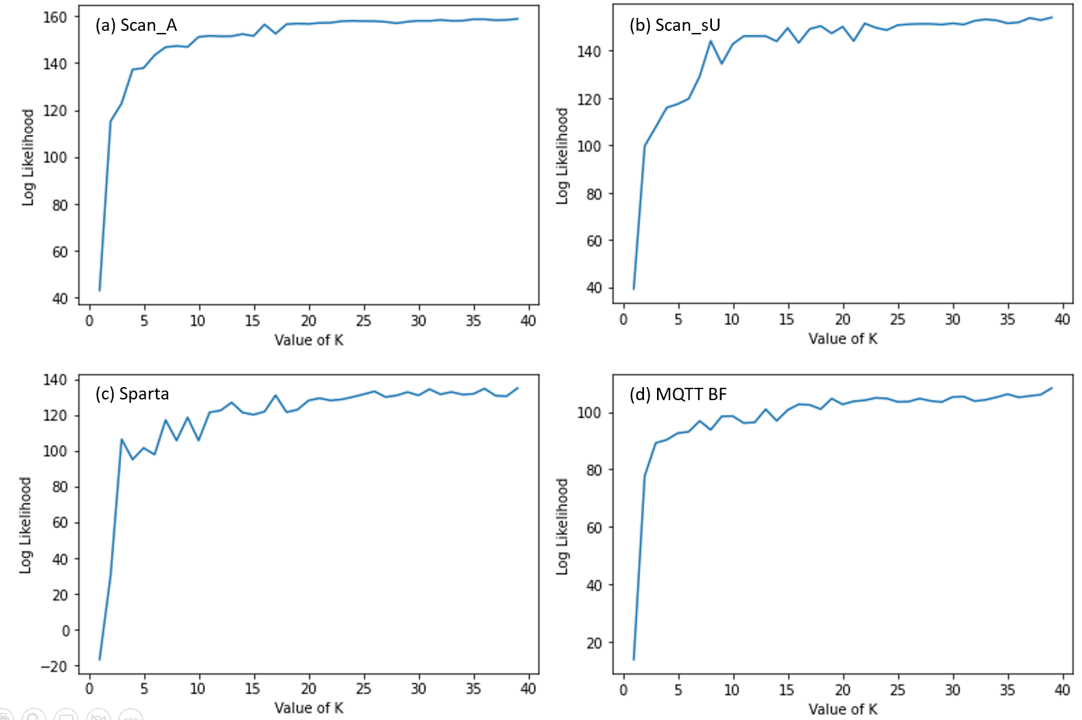}\\
        \end{tabular}
         \vspace{-5pt}
        \caption{Screeplots of the log-likelihood scores for data in four nodes.}\label{fig:screeK}
    \end{center}
     \vspace{-10pt}
\end{figure}

Based on the scree-plots of the log-likelihood score shown in figure~\ref{fig:screeK}, we chose $K=15$ in GMM when estimating the local data density for all nodes. We then calculated the binary classifiers within each node based on the remaining 28 bidirectional-based features (we refer to \cite{article1} for the detailed description of each feature). We applied the logistic regression (LR), the random forests (RF), and the support vector machine with RBF kernel (SVM) to build the local binary classifiers. Again, $75\%$ of the data within each local node were randomly picked as the training data for obtaining the local binary classifiers. The remained data for all nodes were pooled together as the multi-class testing data.
We compared the testing precisions and recalls for the multi-class classifiers obtained through the proposed ensemble approach and the full-data approach. 
The results are listed in table~\ref{tab:mqtt}. 
The values outside the brackets are the mean precision and recall based on 50 replicates, and the values within the brackets are the corresponding standard deviations.

From the table, we observe that the ensemble approach performs similarly to the full-data approach for all three classification models. Such an observation validates the feasibility of the ensemble approach, which can indeed obtain the global multi-class classifier while data in each node is incomplete.

%%%%  Combined table 
\begin{table}[H]
%\begin{center}
\caption{Mean (SD) of the testing precision and recall for each type of attack under the ensemble approach and full-data approach with local classification approaches $(d=28)$.}
\centering
\resizebox{0.8\textwidth}{!}{%
    \begin{tabular}{r|lllll}
        \hline
        \multicolumn{6}{c}{\textbf{Ensemble Multi-node Multi-class Classification}}\\ \hline
          &  Benign & Sparta & Scan-sU & Scan-A & MQTT-BF  \\ \hline 
          & \multicolumn{5}{c}{Precision}\\
         LR & $99.97\%(.008\%)$ & $100\%(0)$& $100\%(0)$ &$100\%(0)$ & $100\%(0)$ \\
         
         RF & $99.97\%(.009\%)$ & $100\%(0)$& $100\%(0)$ &$100\%(0)$ & $100\%(0)$ \\
         
         SVM & $99.97\%$ ($.01\%$)& $100\%(0)$& $100\%(0)$ &$100\%(0)$ & $100\%(0)$ \\ 
         kNN & $99.97\%(.009\%)$ & $100\%(0)$& $100\%(0)$ &$100\%(0)$ & $100\%(0)$ \\        \hline
         & \multicolumn{5}{c}{Recall}\\
         LR & $100\%(0)$  & $99.90\%(.05\%)$ & $ 99.99\%(.01\%)$ &  $99.98(.01\%)$ & $99.96\%(.02\%)$ \\
         
         RF & $100\%(0)$  & $99.89\%(.05\%)$ & $ 99.99\%(.009\%)$ &  $99.98\%(.01\%)$ & $99.98\%(.03\%)$ \\
         
         SVM & $100\%(0)$  & $99.90\%(.06\%)$ & $ 99.99\%(.01\%)$ &  $99.98\%(.01\%)$ & $99.97\%(.01\%)$ \\ 
         kNN & $100\% (0)$ & $99.90\%(.07\%)$& $99.99\%(.01\%)$& $99.98\%(.01\%)$& $99.97\%(.01\%)$ \\ \hline
         \hline
         
         \multicolumn{6}{c}{\textbf{Full Data}}\\ \hline
          & \multicolumn{5}{c}{Precision}\\
         LR & $99.99\%(.002\%)$  & $100\%(0\%)$ & $99.92\%(.03\%)$ &  $ 100\%(0\%)$ & $99.53\%(.13\%)$ \\ 
         RF & $100\%(0\%)$  & $100\%(0\%)$ & $ 99.99\%(.01\%)$ &  $100\%(0\%)$ & $99.99\%(.02\%)$ \\

         SVM & $100\%(0\%)$ & $100\%(0\%)$ & $100\%(0\%)$ &  $99.99\%(.008\%)$ & $99.41\%(.12\%)$ \\
         kNN & $99.99\%(.004\%)$  & $100\%(0\%)$ & $100\%(0\%)$ &  $100\%(0\%)$ & $99.99\%(.01\%)$ \\
        \hline
          & \multicolumn{5}{c}{Recall}\\
         LR & $100\%(0\%)$  & $99.99\%(.01\%)$ & $99.72\%(.08\%)$ &  $99.97\%(.02\%)$ & $99.88\%(..05\%)$ \\ 
         RF & $100\%(0\%)$  & $100\%(0\%)$ & $ 99.99\%(.01\%)$ &  $99.99\%(.008\%)$ & $100\%(0\%)$ \\
         SVM & $99.99\%(.002\%)$  & $100\%(0\%)$ & $ 99.63\%(.08\%)$ & $100\%(0\%)$ & $100\%(0\%)$ \\
         kNN & $99.99\%(.002\%)$  & $99.99\%(.01\%)$ & $100\%(0\%)$ &  $99.97\%(.01\%)$  & $100\%(0\%)$ \\
        \hline
        \end{tabular}
}
%\end{center}
\label{tab:mqtt}
\end{table}

\subsection{Classification with Dimension Reduction}
In the previous experiment, we applied the proposed approach on the original dataset with a dimension of 28. However, based on the studies on the feature subspaces for each type of attack, we observe that the attacks are specified on only four dimensions. To show the utility of the proposed approach, we conducted a dimension reduction via Principal Component Analysis (PCA), and reduced the data dimension from 28 to four. %Thus, we first do the Principal Component Analysis, reducing the dimension of the data to 4, and then repeat the process. In this section, we regard the original dataset as the result after PCA. 
Again, four binary classification algorithm, namely Logistic Regression(LR), Random Forest(RF), k Nearest Neighborhood(kNN), Support Vector Machine (SVM) were adopted for the task after dimension reduction, and the results are listed in table~\ref{tab:mqttPCA}. 

\begin{table}[H]
%\begin{center}
\caption{Mean (SD) of the testing precision and recall for each type of attack under the ensemble approach and full-data approach with local classification approaches after PCA $(d=4)$.}
\centering
\resizebox{0.8\textwidth}{!}{%
    \begin{tabular}{r|lllll}
        \hline
        \multicolumn{6}{c}{\textbf{Ensemble Multi-node Multi-class Classification}}\\ \hline
          &  Benign & Sparta & Scan-sU & Scan-A & MQTT-BF  \\ \hline 
          & \multicolumn{5}{c}{Precision}\\
         LR & $98.76\%(.37\%)$ & $100\%(0\%)$ & $99.87\%(.21\%)$ & $99.72\%(.06\%)$ & $97.06\%(.20\%)$ \\
         
        RF & $99.99\%(.002\%)$  & $100\%(0\%)$ & $ 99.99\%(.01\%)$ &  $99.99\%(.006\%)$ & $97.07\%(.20\%)$ \\
         
         SVM & $99.90\%(.30\%)$ & $100\%(0\%)$& $99.99\%(.01\%)$ &$99.71\%(.06\%$ & $97.06\%(.20\%)$ \\ 
         
         kNN & $99.97\%(.01\%)$ & $100\% (0\%)$ & $99.99\%(.01\%)$& $99.99\%(.007\%)$& $97.07\%(.002\%)$ \\        \hline
         & \multicolumn{5}{c}{Recall}\\
         LR & $99.49\%(.06\%)$  & $100\%(0\%)$ & $ 99.67\%(.67\%)$ &  $94.38\%(1.8\%)$ & $100\%(0\%)$ \\
         
         RF & $99.58\%(.03\%)$  & $100\%(0\%)$ & $ 99.99\%(.003\%)$ &  $99.99\%(.007\%)$ & $99.99\%(.01\%)$ \\

         SVM & $99.52\%(.03\%)$  & $100\%(0\%)$ & $ 99.93\%(.02\%)$ &  $99.59\%(1.4\%)$ & $99.99\%(.01\%)$ \\ 
         kNN & $99.58\% (.03\%)$ & $100\%(0)$& $99.94\%(.03\%)$& $99.95\%(.03\%)$& $99.99\%(.006\%)$ \\ \hline
         \hline
         
         \multicolumn{6}{c}{\textbf{Full Data}}\\ \hline
          & \multicolumn{5}{c}{Precision}\\
         LR & $99.43\%(.07\%)$  & $100\%(0\%)$ & $79.39\%(.58\%)$ &  $ 85.22\%(4.3\%)$ & $97.05\%(.21\%)$ \\ 
         RF & $99.75\%(.03\%)$  & $100\%(0\%)$ & $ 100\%(0\%)$ &  $100\%(0\%)$ & $98.01\%(.16\%)$ \\
         
         SVM & - & - & - & -& -\\

         kNN & $99.97\%(.02\%)$  & $100\%(0\%)$ & $100\%(0\%)$ &  $100\%(0\%)$ & $97.09\%(.20\%)$ \\
        \hline
          & \multicolumn{5}{c}{Recall}\\
         LR & $92.75\%(.78\%)$  & $100\%(0)$ & $99.71\%(.25\%)$  & $96.84\%(.26\%)$&$99.98\%(.02\%)$ \\ 
         RF & $99.71\%(.02\%)$  & $100\%(0\%)$ & $ 100\%(0\%)$ &  $99.99\%(.004\%)$ & $98.29\%(.23\%)$ \\
         
                  SVM & - & - & - & -& -\\

         kNN & $99.58\%(.03\%)$  & $99.98\%(.02\%)$ & $99.97\%(.03\%)$ &  $99.97\%(.02\%)$  & $99.88\%(.12\%)$ \\
        \hline

        \end{tabular}
}
%\end{center}
\label{tab:mqttPCA}
\end{table}

%\begin{remark}
Since the Support Vector Machine method for full-data is super time-consuming compared with other methods, we did not report the results for SVM under the full-data case. %Compared to the full-data approaches, the proposed approach save lots of time, and it is proper to deal with such classifiers who require large time complexity when doing multi-class classification for many classes.
%\end{remark}
%\begin{remark}
According to the results, when applying Logistic Regression for classification, the proposed multi-node multi-class classification ensemble approach has a higher Precision compared with the full-data approach. One possible reason is that after processing data via PCA, data may become much more dense, the linear classifier (LR) may not be suitable for the multi-class classification. However, it may still be suitable for the simple binary classification within a single node. In addition, GMM may also results in a more accurate density estimation for a dense data with less dimension than for a high-dimensional data. As a result, the proposed approach performs better than the full-data approach with LR.
%\end{remark}

\section{Conclusion and discussion}
In this paper, we study the federated learning framework for multi-class classification. We focus on the cases when local nodes only contain a certain part of the classes. By filling in the missing information through gathering the parameters from local binary classifiers and local densities, we develop a multi-node multi-class classification ensemble approach. We validate the feasibility of the proposed approach both theoretically and practically. 
We also discuss the impact of data balancing. Actually, there are some existing weighted binary classification approaches for imbalanced data \cite{ganganwar2012overview, krawczyk2016learning}. When we adopt such weighted approaches to the local classifiers, the proposed ensemble approach can be potentially extended to a more general approach for imbalanced data. 
Furthermore, the proposed approach has three interesting merits in aspects of communication complexity, privacy, and generality. Firstly, it's one-shot in terms of communication complexity \cite{guha2019one, salehkaleybar2021one}, in that it only requires one communication between the server and each client. In contrast, conventional federated learning algorithms require iterative communications, which is time-consuming and demands that different clients must be synchronized. Thanks to the density estimator, our algorithm is further free from the need for a global public dataset or synthetic data generation/distillation, which are required by existing one-shot federated learning approaches \cite{zhou2020distilled, kasturi2020fusion}.
Secondly, the proposed approach does not transmit gradient information, which makes it intrinsically immune to common attack strategies in decentralized machine learning such as the gradient inversion attack \cite{geiping2020inverting}. 
Thirdly, the proposed approach is general enough that all sorts of classifiers and density estimators can be incorporated to replace the currently used binary classifiers and GMM models. This allows one to borrow cutting-edge techniques from deep learning or even quantum machine learning communities, such as more advanced neural networks or quantum circuits as classifiers and modern generative models as density estimators \cite{goodfellow2016deep, biamonte2017quantum}.

% or


% \appendices

% % you can choose not to have a title for an appendix
% % if you want by leaving the argument blank
% \section{}

% \section*{Acknowledgment}


\appendix
\section*{Appendix}

\section{Proof of the main result} \label{app:proof}
Let $w_j = \frac{f_{\x}^{(j)}p_N^{(j)}}{\sum_{j=1}^Jf_{\x}^{(j)}p_N^{(j)}}$. 
We first show the feasibility of \eqref{eq:globalC}.
\begin{equation*}
\begin{split}
    c^{(m)}(\x) = & P(y = m|\x)\\
    = &\sum_{j=1}^N P(y = m, \mathrm{data}\in \mathrm{Node}_j|\x) \\
    = &\sum_{j=1}^N \frac{P(y = m, \mathrm{data}\in \mathrm{Node}_j, \x)}{f_{\x}} \\
    = &\sum_{j=1}^N \frac{P(y = m, \mathrm{data}\in \mathrm{Node}_j, \x)}{P(\mathrm{data}\in \mathrm{Node}_j, \x)}\times\frac{P(\mathrm{data}\in \mathrm{Node}_j, \x)}{f_{\x}} \\
    = &\sum_{j=1}^N c_j^{(m)}(\x)\times\frac{f_{\x}^{(j)}p_N^{(j)}}{f_{\x}} \\
    = &\sum_{j=1}^N c_j^{(m)}(\x)\times\frac{f_{\x}^{(j)}p_N^{(j)}}{\sum_{j=1}^N f_{\x}^{(j)}p_N^{(j)}} \\
    = &\sum_{j=1}^J1(m\in\mathcal{M}_j)w_jc_j^{(m)}(\x).
\end{split}
\end{equation*}

We then prove that when $w_j\ge \left(\hat{c}_j^{(m)}\right)^{J-1}/J$, $\hat{c}^{(m)}(\x)= \sum_{j=1}^J1(m\in\mathcal{M}_j)w_j\hat{c}_j^{(m)}(\x)$ minimizes loss function~\eqref{eq:multiclass}.

Consider the term $\sum_m\log c^{(m)}(\x_{ij})$ in \eqref{eq:multiclass}. Plug-in \eqref{eq:globalC}, we have
\begin{equation}\label{eq1}
\sum_m\log \hat{c}^{(m)}(\x_{ij}) = \sum_m\log \sum_{j=1}^J1(m\in\mathcal{M}_j)w_j\hat{c}_j^{(m)}(\x_{ij})= \sum_m\log\sum_{j=1}^Jw_j\hat{c}_j^{(m)}(\x_{ij}).
\end{equation}
By Jensen's inequality, we have
\begin{equation}\label{eq2}
\log\sum_{j=1}^Jw_j\hat{c}_j^{(m)}(\x_{ij}) \ge \frac{1}{J}\sum_{j=1}^J\log Jw_j\hat{c}_j^{(m)}(\x_{ij}).
\end{equation}
When $w_j\ge \left(\hat{c}_j^{(m)}\right)^{J-1}/J$, then
\begin{equation}\label{eq3}
\frac{1}{J}\sum_{j=1}^J\log Jw_j\hat{c}_j^{(m)}(\x_{ij}) = \sum_{j=1}^J\log\left(Jw_j\hat{c}_j^{(m)}(\x_{ij})\right)^{1/J} \ge \sum_{j=1}^J\log\hat{c}_j^{(m)}(\x_{ij}).
\end{equation}
Combining \eqref{eq1}, \eqref{eq2} and \eqref{eq3}, we have
\begin{equation}\label{eq4}
-\sum_i\sum_m1(y_{ij}=m)\log \hat{c}^{(m)}(\x_{ij}) \le -\sum_i\sum_m\sum_{j=1}^J1(y_{ij}=m)\log\hat{c}_j^{(m)}(\x_{ij}).
\end{equation}
What's more, $\sum_m\hat{c}^{(m)} = \sum_jw_j\sum_m1(m\in\mathcal{M}_j\hat{c}_j^{(m)}) = 1$.
On the other hand, since $\hat{c}_j^{(m)}$ minimizes \eqref{eq:single} for $j = 1, \ldots, J$, it is obvious that $\sum_j\left\{-\sum_i\sum_m1(y_{ij}=m)\log\hat{c}_j^{(m)}(\x_{ij})\right\}$ reaches its minimum. Together with \eqref{eq4}, we show that the ensemble estimation $\hat{c}^{(m)}$ minimizes the global loss function \eqref{eq:multiclass}.

\section{Application on MNIST dataset} \label{app:mnist}

MNIST \cite{lecun1998mnist} is a popular image dataset with 60,000 samples that contains 10 classes of handwritten digits from 0 to 9. In such a dataset, there is no obvious justification for the particular ``star'' structure. We thus additionally consider the ``ring'' and ``fully-connected'' structures. 
We randomly pick $75\%$ of the data within each node to train the local binary classifiers. We then combine the remained data ($25\%$ for each node) together as the multi-class test data.
We use the random forest (RF) as the binary classifiers. 
We compare the test precision and recall of the multi-class classifiers obtained through the proposed ensemble approach and the full-data approach. Here the ``full-data approach'' refers to the classifier trained with the pooled data. The results averaged over 50 replicates are listed in table~\ref{tab:mnist}.
The values outside the brackets are the mean precision and recall based, and the values within the brackets are the corresponding standard deviations. 
From the table, we observe that the ensemble approach under the ``fully-connected'' structure is slightly better than the ``ring'' structure. Such an observation is expected since the ``fully-connected'' structure contains more mutual information among classes. We also observe that the ensemble approach performs similarly to the full-data approach. Such an observation validates the feasibility and effectiveness of the federated approach.

\begin{table}[!ht]
\caption{Mean (SD) of the testing precision and recall for each digit.}
\centering
\resizebox{0.5\textwidth}{!}{%
\begin{tabular}{r|lll}
\hline
              & Ring  & Fully-connected & Full-data \\ \hline
Digit         & \multicolumn{3}{c}{Precision}\\
0   & $96.91\%(3.19\%)$ & $97.19\% (2.79\%)$   &  $98.92\%(1.32\%)$         \\ %\hline
1   &   $94.21\%(4.70\%)$         & $96.62\%(2.38\%)$         &    $97.17\%(1.27\%)$          \\ %\hline
2 &   $94.11\%(3.02\%)$         & $96.80\% (3.31\%)$          &   $97.67\%(2.17\%)$        \\ %\hline
3   &   $93.13\%(2.03\%)$         & $96.80\%(3.14\%)$         &    $97.41\%(2.55\%)$          \\ %\hline
4   &   $91.12\%(3.72\%)$         & $92.97\%(2.37\%)$         &    $92.92\%(2.35\%)$          \\ %\hline
5 &   $95.57\%(4.21\%)$         & $96.45\% (4.54\%)$          &   $97.17\%(2.47\%)$        \\ %\hline
6   &   $97.74\%(2.26\%)$         & $97.35\%(1.90\%)$         &    $98.61\%(1.27\%)$          \\ %\hline
7 &   $98.21\%(2.04\%)$         & $99.05\% (1.47\%)$          &   $99.42\%(1.16\%)$        \\ %\hline
8 &   $95.89\%(4.10\%)$         & $96.59\% (4.37\%)$          &   $95.57\%(3.92\%)$        \\ %\hline
9 &   $92.17\%(3.63\%)$         & $92.57\% (3.65\%)$          &   $93.31\%(2.85\%)$        \\ \hline
              & \multicolumn{3}{c}{Recall}\\
0   & $93.41\%(2.30\%)$         & $94.58\% (3.40\%)$   &  $94.46\%(2.57\%)$         \\ %\hline
1  &   $97.53\%(1.93\%)$         & $97.99\%(2.29\%)$         &    $97.15\%(2.17\%)$          \\ %\hline
2 &   $97.59\%(2.64\%)$         & $98.64\% (2.24\%)$          &   $98.23\%(1.72\%)$        \\ %\hline
3   & $96.23\%(2.67\%)$         & $96.62\% (2.95\%)$   &  $96.63\%(2.52\%)$         \\ %\hline
4  &   $96.20\%(2.64\%)$         & $97.94\%(2.24\%)$         &    $97.67\%(2.51\%)$          \\ %\hline
5 &   $96.52\%(1.64\%)$         & $96.51\% (2.36\%)$          &   $96.99\%(2.92\%)$        \\ %\hline
6   & $99.78\%(0.67\%)$         & $99.77\% (0.70\%)$   &  $99.77\%(0.68\%)$         \\ %\hline
7  &   $93.48\%(4.54\%)$         & $95.35\%(3.30\%)$         &    $94.67\%(2.52\%)$          \\ %\hline
8 &   $94.97\%(3.82\%)$         & $95.35\% (3.37\%)$          &   $96.13\%(3.41\%)$        \\ %\hline
9 &   $96.64\%(4.00\%)$         & $96.91\% (3.10\%)$          &   $97.01\%(3.46\%)$        \\ \hline
\end{tabular}%
}
\label{tab:mnist}
\end{table}

\section*{Acknowledgements}
The authors would like to acknowledge support for this project from the National Key R\&D Program of China (No. 2021YFA1001300).

%\ifCLASSOPTIONcaptionsoff
  \newpage
%\fi

\bibliographystyle{unsrt}
\bibliography{ref}

\end{document}